\documentclass[aps,prx,onecolumn,superscriptaddress,longbibliography]{revtex4-1}
\usepackage{titlesec}
\usepackage{amsmath}  
\usepackage{amsfonts} 
\usepackage{graphicx}
\usepackage{amssymb}
\usepackage{hyperref} 
\hypersetup{colorlinks=true,urlcolor=blue,citecolor=blue,linkcolor=blue}
\usepackage{cleveref}
\usepackage{etoolbox}
\AtBeginEnvironment{quote}{\singlespacing\small}

\DeclareMathOperator{\dd}{d\!}

\begin{document}

\title{Demystifying the Lagrangians of Special Relativity}

\author{Gerd Wagner}
\email{gerdhwagner@t-online.de} 
\affiliation{Mayener Str. 131, 56070 Koblenz, Germany} 

\author{Matthew W. Guthrie}
\email{matthew.guthrie@uconn.edu}
\affiliation{Department of Physics, University of Connecticut, Storrs, CT 06269}


\date{\today}


\begin{abstract}

Special relativity beyond its basic treatment can be inaccessible, in particular because introductory physics courses typically view special relativity as decontextualized from the rest of physics. 
We seek to place special relativity back in its physics context, and to make the subject approachable.
The Lagrangian formulation of special relativity follows logically by combining the Lagrangian approach to mechanics and the postulates of special relativity.
In this paper, we derive and explicate some of the most important results of how the Lagrangian formalism and Lagrangians themselves behave in the context of special relativity.
We derive two foundations of special relativity: the invariance of any spacetime interval, and the Lorentz transformation. 
We then develop the Lagrangian formulation of relativistic particle dynamics, including the transformation law of the electromagnetic potentials, the Lagrangian of a relativistic free particle, and Einstein's mass-energy equivalence law ($E=mc^2$).
We include a discussion of relativistic field Lagrangians and their transformation properties, showing that the Lagrangians and the equations of motion for the electric and magnetic fields are indeed invariant under Lorentz transformations.

\end{abstract}

\maketitle

\section{Introduction}

\subsection{Motivation}

Lagrangians link the relationship between equations of motion and coordinate systems.
This is particularly important for special relativity, the foundations of which lie in considering the laws of physics within different coordinate systems (i.e., inertial reference frames).
The purpose of this paper is to explicate the physically convenient fact that once the transformation rules of the Lagrangian formalism are derived, major results of relativistic electrodynamics can be found by simply applying these rules (this is a common omission from special relativity courses).

In the current paper, we apply the results of our previous work on the Lagrangian formulations of classical mechanics and field theory~\cite{WagnerGuthrie, WagnerGuthrieClassicalField} to the field of special relativity.
Analogous to our discussion of the Lagrangian formulation of classical mechanics~\cite{WagnerGuthrie}, in which we discussed the Lagrangian formalism for classical particle mechanics, we will show that the Lagrangian formalism holds significant advantages for analyses in relativistic particle mechanics, as well.

Nonetheless, in the Lagrangian formulation of relativistic particle dynamics, we encountered limitations to the allowable coordinate transformations (i.e., only transformations of the spatial and time coordinates are allowed which are invertible in the time coordinate).
However, no such limitations occur in the Lagrangian formalism for relativistic fields.
Relativistic field Lagrangians, as we will show, always have well-behaved transformation properties; the Euler-Lagrange equations are invariant under arbitrary transformations of spatial and time coordinates.

We apply these results to derive several facets of the theory of special relativity. 
First and foremost, we show that the electromagnetic potentials form a 4-vector.
This result follows from our discussion of the relativistic particle Lagrangian and from Einstein's postulates of relativity applied to the Lorentz force.
Based on this result, we can then show that Maxwell's field equations do not change under Lorentz transformations.

\subsection{Review}

One major departure we take from our previous work~\cite{WagnerGuthrie, WagnerGuthrieClassicalField} stems from how those papers required minimal prerequisites: namely, Newton's second law and the classical formulation of Maxwell's equations.
In the interest of making this paper similarly self-contained, we must derive several well-known results from special relativity. 
Those results are also shown in the following works:
\begin{itemize}
    \item Similar to Landau~\cite{LandauInterval}, we show that spatial and time coordinates of any particle form is a 4-vector.
    \item Similar to Ref.~\cite{idema2018mechanics}, we derive the Lorentz transformations.
    \item The facts we cite about the spacetime position 4-vector, the 4-velocity and the rules that 4-vectors obey are a small subset of what is discussed in Ref.~\cite{mcmahon2005relativity}.
    \item The derivation of energy conservation we present can be found in Refs.~\cite{classicalTimeTranslationEnergyConservation, GoldsteinConservation}.
    \item A recent paper which discusses the Lagrangian of the Lorentz force and which arrives at the same results as we do is in Ref.~\cite{davis2019energy}.
    \item We derive mass-energy equivalence as in Refs.~\cite{mcmahon2005relativity, youtube_SusskindSpecialRelativity3}.
    \item Our discussion of gauge of the particle Lagrangian and of the electromagnetic potentials can be found in Refs.~\cite{Zinn-Justin:2008, srednicki_2007}.
    \item Our proof that the electromagnetic charge and current densities form a 4-vector is also contained in Ref.~\cite{weinberg2005quantum}.
\end{itemize}

The presentation of those results is in service of our treatment of the Lagrangian formulation of special relativity. 
One recent treatise of special relativity in which the discussion of prerequisites comes close to ours is found in Ref.~\cite{zakamska2018theory}.
A critical difference is that the Lagrangian formalism is the focus of our paper, which is not found in Zakamaska's treatment.
A less technical treatise of special relativity can be found in Ref.~\cite{mansson2009understanding}, which likewise does not discuss the relativistic Lagrangian formalism.
Chanda and Guha~\cite{Chanda_2018} discuss relativistic particle Lagrangians in detail, however that discussion does not consider the transformation properties of the Lagrangian and the Euler-Lagrange equations. 
These properties are crucial for motivating our development of the relativistic Lagrangian formalism.

We begin our development of these properties by showing the ways in which space and time coordinates behave when considering the postulates of special relativity, and the consequences thereof.
That discussion begins in Sec.~\ref{foundations}.

\section{Two foundations of special relativity} \label{foundations}

\subsection{Invariance of the spacetime interval $\Delta s$ } \label{sectionInvarianceSpaceTime}

Following the example of Ref.~\cite{LandauInterval}, we consider two inertial reference frames (IRFs) $T$ and $\bar{T}$, with times $t$ and $\bar{t}$, as well as Cartesian coordinates $x$, $y$, $z$ and $\bar{x}$, $\bar{y}$, $\bar{z}$.
IRFs are reference frames in which a body with zero net force acting upon it does not accelerate.
The possible transformations between IRFs are translations in space and time, rotations in space, and motion with constant velocity.

Imagine that, within $T$ and $\bar{T}$, we observe a particle and a light pulse.
Assume the light pulse travels in one direction with velocity $c$ such that it has nearly as well defined position as the particle.

The empirical fact which started the theory of special relativity is that any change in the coordinates and time ($\Delta x, \Delta y, \Delta z, \Delta t$) in the first reference frame corresponds to changes in coordinates and times for the second reference frame
\begin{equation} \label{constantsSpeedOfLight}
0 = c^2\Delta t^2 - \Delta x^2 - \Delta y^2 - \Delta z^2 = c^2\Delta \bar{t}^2 - \Delta \bar{x}^2 - \Delta \bar{y}^2 - \Delta \bar{z}^2;
\end{equation}
that is, any observation of the light pulse's coordinates holds in the two inertial reference frames $T$ and $\bar{T}$ \footnote{This is the usual mathematical way to state that the speed of light is the same in all IRFs.
The fact that the speed of light is the same in all IRFs is also known as the second postulate of special relativity.
There are altogether two postulates of special relativity.
The two-postulate basis for special relativity is the one historically used by Einstein, and it remains the starting point today.
The first postulate, which is also called ``principle of relativity'' and postulates that the laws of physics are the same in all IRFs, will be discussed in sections \ref{relativisticParticleEulerLagrangeInvariance}, \ref{sectionHeuristicFirstPostulate}, and \ref{sectionEinsteinsOriginalFirstPostulate} of this paper.}.

If $T$ and $\bar{T}$ are relative to each other at rest, the equality
\begin{equation}
c^2\Delta t^2 - \Delta x^2 - \Delta y^2 - \Delta z^2 = c^2\Delta \bar{t}^2 - \Delta \bar{x}^2 - \Delta \bar{y}^2 - \Delta \bar{z}^2
\end{equation}
is true for the particle, too.
Without relative motion, the only transformations left are translations in space and time, and rotations in space.
For these transformations the even more restrictive relations
\begin{equation} \label{noVelocityTransform}
\Delta t = \Delta \bar{t} \; \text{and} \; \Delta x^2 + \Delta y^2 + \Delta z^2 = \Delta \bar{x}^2 + \Delta \bar{y}^2 + \Delta \bar{z}^2
\end{equation}
hold. 
If these relations did not hold when $T$ and $\bar{T}$ are relative to each other at rest, this would be a strong hint that space is not homogeneous and isotropic.
The issues that lead to special relativity arise only when $T$ and $\bar{T}$ are moving relative to each other.\\

The question we now aim to answer is: Is it true that for any transformation connecting the two IRFs the equation
\begin{equation} \label{invarianceSpaceTimeInterv}
    c^2\Delta t^2 - \Delta x^2 - \Delta y^2 - \Delta z^2 = c^2\Delta \bar{{t}}^2 - \Delta \bar{x}^2 - \Delta \bar{y}^2 - \Delta \bar{z}^2
\end{equation}
holds for the particle's coordinates in $T$ and $\bar{T}$, as well?

As we reasoned above, the only transformations this question is not already answered for are those which include a constant relative velocity $\vec{v}$ between $T$ and $\bar{T}$.
As spatial rotations have no effect on
\begin{equation}
    \Delta s^2 := c^2\Delta t^2 - \Delta x^2 - \Delta y^2 - \Delta z^2
\end{equation}
and
\begin{equation}
\Delta \bar{s}^2 := c^2\Delta \bar{t}^2 - \Delta \bar{x}^2 - \Delta \bar{y}^2 - \Delta \bar{z}^2,
\end{equation}
the direction of $\vec{v}$ cannot have an effect either. As such, only the magnitude of the velocity $|\vec{v}|$ could be responsible for $\Delta s^2 \neq \Delta \bar{s}^2 $.
Thus, the possible relation between $\Delta \bar{s}$ and $\Delta s$ can be expressed by a family of functions $F$ parameterized by $|\vec{v}|$ such that
\begin{equation}
    \Delta \bar{s} = F_{|\vec{v}|} (\Delta s).
\end{equation}
Below we will show that $F$ cannot depend on $|\vec{v}|$ at all.
If so, then $F$ relates $\Delta \bar{s}$ and $\Delta s$ the same way no matter what the transformation between $T$ and $\bar{T}$ is.
As a result, we can use any special case of a transformation to determine $F$.
Because any of Eqs. \eqref{noVelocityTransform} and \eqref{constantsSpeedOfLight} lead to
\begin{equation}
    \Delta \bar{s} = \Delta s,
\end{equation}
$F$ must be the identity function, and Eq. \eqref{invarianceSpaceTimeInterv} must hold for the particle, too.
\\

To prove that $F$ cannot depend on $|\vec{v}|$, we consider three IRFs $T_1$, $T_2$, $T_3$ with constant relative velocities

\begin{align*}
    &\vec{v}_{12} \; \text{between} \; T_1 \; \text{and} \; T_2 \\
    &\vec{v}_{23} \; \text{between} \; T_2 \; \text{and} \; T_3 \\
    &\vec{v}_{13} \; \text{between} \; T_1 \; \text{and} \; T_3. \\
\end{align*}
The equations for the relations between $\Delta s_1$, $\Delta s_2$, and $\Delta s_3$ would then read
\begin{align}
\Delta s_2 &= F_{|\vec{v}_{12}|}(\Delta s_1) \label{Fofv1} \\
\Delta s_3 &= F_{|\vec{v}_{13}|}(\Delta s_1) \label{Fofv2} \\
\Delta s_3 &= F_{|\vec{v}_{23}|}(\Delta s_2) \label{Fofv3}.
\end{align}
Substituting Eq. \eqref{Fofv1} into Eq. \eqref{Fofv3} leads to $\Delta s_3 = F_{|\vec{v}_{23}|}\left(F_{|\vec{v}_{12}|} (\Delta s_1)\right)$ which, with Eq. \eqref{Fofv2}, leads to the identity
\begin{equation} \label{FRelation}
F_{|\vec{v}_{13}|}(\Delta s_1) = F_{|\vec{v}_{23}|} \left( F_{|\vec{v}_{12}|}(\Delta s_1) \right).
\end{equation}
At the same time, the vector equation $\vec{v}_{13} = \vec{v}_{12} + \vec{v}_{23}$ must hold. For the absolute values, this means
\begin{equation}
    |\vec{v}_{13}| = \sqrt{|\vec{v}_{12}|^2 + |\vec{v}_{23}|^2 + 2 |\vec{v}_{12}||\vec{v}_{23}| \cos(\alpha)},
\end{equation}
where $\alpha$ is the angle between $\vec{v}_{12}$ and $\vec{v}_{23}$.
Consequently, the left hand side of Eq. \eqref{FRelation} would depend on $\alpha$, but the right hand side would not.
Because of this contradiction, $F$ cannot depend on the relative velocity between IRFs.

\subsection{The Lorentz transformation} \label{sectionLorentzTransformation}

\subsubsection{Non-relativistic setup} \label{setup}
We observe a particle from within two IRFs $T$ and $\bar{T}$.
The axes of the frames point in the same direction, and $\bar{T}$ moves with velocity $v$ along $T$'s $x$-axis such that their classical Galilean transformation is given by:
\begin{equation} \label{Galilei}
\left(\begin{array}{c}
\bar{t}
\\
\bar{x}
\end{array} \right)
=
\begin{pmatrix}
1 & 0
\\
-v & 1
\end{pmatrix}
\left(\begin{array}{c}
t
\\
x
\end{array} \right),
\; \bar{y}=y ,\; \bar{z}=z,
\end{equation}
where $t$, $x$, $y$, $z$ are the coordinates of the particle in $T$, and $\bar{t}$, $\bar{x}$, $\bar{y}$, $\bar{z}$ are the coordinates of the particle in $\bar{T}$.
As evident in these equations, we chose $t=\bar{t}=0$ to be the time when the coordinate frames are on top of one another.

\subsubsection{Ansatz for a transformation compatible with invariance of the spacetime interval}

Because at $t=\bar{t}=0$, the two reference frames $T$ and $\bar{T}$ are on top of each other, Eq. \eqref{invarianceSpaceTimeInterv} becomes
\begin{equation} \label{spaceTimeElement}
c^2\bar{t}^2-\bar{x}^2-\bar{y}^2-\bar{z}^2 = c^2t^2-x^2-y^2-z^2.
\end{equation}
As an ansatz for a coordinate transformation between the two reference frames which is consistent with Eq. \eqref{spaceTimeElement}, we write:
\begin{align} \label{ansatz}
\left(\begin{array}{c}
c\bar{t}
\\
\bar{x}
\end{array} \right)
&=
\begin{pmatrix}
a & d
\\
b & e
\end{pmatrix}
\left(\begin{array}{c}
ct
\\
x
\end{array} \right) \\
\bar{y} &= y \nonumber \\
\bar{z} &= z, \nonumber
\end{align}
which we utilize in the following section.

\subsubsection{Finding the matrix elements of the ansatz}

Because $\bar{y}=y$, $\bar{z}=z$, Eq. \eqref{spaceTimeElement} and ansatz \eqref{ansatz} result in the following equation
\begin{equation} \label{matrixEquationLorentzTransform}
\left[
\begin{pmatrix}
a & d
\\
b & e
\end{pmatrix}
\left(\begin{array}{c}
ct
\\
x
\end{array} \right)
\right]^\mathsf{T}
\begin{pmatrix}
1 & 0
\\
0 & -1
\end{pmatrix}
\left[
\begin{pmatrix}
a & d
\\
b & e
\end{pmatrix}
\left(\begin{array}{c}
ct
\\
x
\end{array} \right)
\right]
=
\left(\begin{array}{c}
ct
\\
x
\end{array} \right)^\mathsf{T}
\begin{pmatrix}
1 & 0
\\
0 & -1
\end{pmatrix}
\left(\begin{array}{c}
ct
\\
x
\end{array} \right).
\end{equation}
Since $t$ and $x$ are arbitrary the only way for this equation to be true is
\begin{equation} \label{minkowski}
\begin{pmatrix}
a & b
\\
d & e
\end{pmatrix}
\begin{pmatrix}
1 & 0
\\
0 & -1
\end{pmatrix}
\begin{pmatrix}
a & d
\\
b & e
\end{pmatrix}
=
\begin{pmatrix}
1 & 0
\\
0 & -1
\end{pmatrix}
\end{equation}

\begin{equation}
\iff
\begin{pmatrix}
a & b
\\
d & e
\end{pmatrix}
\begin{pmatrix}
a & d
\\
-b & -e
\end{pmatrix}
=
\begin{pmatrix}
1 & 0
\\
0 & -1
\end{pmatrix}
\end{equation}

\begin{equation}
\iff
\begin{pmatrix}
a^2-b^2 & ad-be
\\
ad-be & d^2-e^2
\end{pmatrix}
=
\begin{pmatrix}
1 & 0
\\
0 & -1
\end{pmatrix}
\end{equation}
These are three equations for four parameters $a,b,d,e$. This means one free parameter will remain.
We are now going to solve these equations in such a way that $a,d,$ and $e$ are expressed through $b$:
\begin{equation}
a(b) = \pm \sqrt{1+b^2}
\end{equation}
\begin{equation}
-1 = d^2-e^2 = \frac{b^2e^2}{a^2} -e^2 = e^2 \left(\frac{b^2}{a^2} - 1 \right)
\end{equation}
\begin{equation}
\iff e^2 = \frac{1}{1-\frac{b^2}{a^2}} = \frac{1}{1-\frac{b^2}{1+b^2}} = \frac{1}{\frac{1+b^2-b^2}{1+b^2}} = 1+b^2
\end{equation}
\begin{equation}
\iff e(b) = \pm \sqrt{1+b^2}
\end{equation}
\begin{equation}
d^2 = e^2 -1 = b^2 \iff d(b) = \pm b
\end{equation}
\begin{equation} \label{matrixExpressedByB}
\implies
\begin{pmatrix}
a & d
\\
b & e
\end{pmatrix}
=
\begin{pmatrix}
\sqrt{1+b^2} & b
\\
b & \sqrt{1+b^2}
\end{pmatrix}.
\end{equation}
In evaluating Eq. \eqref{matrixExpressedByB} we chose the positive solutions only.
This can be rectified by checking that the matrix \eqref{matrixExpressedByB} solves Eq. \eqref{minkowski}. The transformation law now reads
\begin{equation}
\left(\begin{array}{c}
c\bar{t}
\\
\bar{x}
\end{array} \right)
=
\begin{pmatrix}
\sqrt{1+b^2} & b
\\
b & \sqrt{1+b^2}
\end{pmatrix}
\left(\begin{array}{c}
ct
\\
x
\end{array} \right)
\;,\; \bar{y}=y \;,\; \bar{z}=z.
\end{equation}
Next we want to move the speed of light $c$ from the vectors to the matrix.
To do so we write the above equation in the following form:
\begin{align}
\begin{pmatrix}
c & 0
\\
0 & 1
\end{pmatrix}
\left(\begin{array}{c}
\bar{t}
\\
\bar{x}
\end{array} \right)
&=
\begin{pmatrix}
\sqrt{1+b^2} & b
\\
b & \sqrt{1+b^2}
\end{pmatrix}
\begin{pmatrix}
c & 0
\\
0 & 1
\end{pmatrix}
\left(\begin{array}{c}
t
\\
x
\end{array} \right) \label{shiftCFromVectorToMatrix} \\
\iff
\left(\begin{array}{c}
\bar{t}
\\
\bar{x}
\end{array} \right)
&=
\begin{pmatrix}
1/c & 0
\\
0 & 1
\end{pmatrix}
\begin{pmatrix}
\sqrt{1+b^2} & b
\\
b & \sqrt{1+b^2}
\end{pmatrix}
\begin{pmatrix}
c & 0
\\
0 & 1
\end{pmatrix}
\left(\begin{array}{c}
t
\\
x
\end{array} \right) \\
\iff
\left(\begin{array}{c}
\bar{t}
\\
\bar{x}
\end{array} \right)
&=
\begin{pmatrix}
\sqrt{1+b^2} /c & b /c
\\
b & \sqrt{1+b^2}
\end{pmatrix}
\begin{pmatrix}
c & 0
\\
0 & 1
\end{pmatrix}
\left(\begin{array}{c}
t
\\
x
\end{array} \right) \\
\iff
\left(\begin{array}{c}
\bar{t}
\\
\bar{x}
\end{array} \right)
&=
\begin{pmatrix}
\sqrt{1+b^2} & b /c
\\
c b  & \sqrt{1+b^2}
\end{pmatrix}
\left(\begin{array}{c}
t
\\
x
\end{array} \right). \label{transformationBasedOnb}
\end{align}

\subsubsection{Calculation of $b$} \label{calcOfB}

We would now like to give a physical interpretation of $b$.
To do so, we consider Eq. \eqref{spaceTimeElement} for the special case where the particle moves with the origin of reference frame $\bar{T}$.

We first use $\bar{y}=y,$ and $\bar{z}=z$ to simplify Eq. \eqref{spaceTimeElement} to
\begin{equation} \label{spaceTimeElementReduced}
c^2\bar{t}^2-\bar{x}^2 = c^2t^2-x^2.
\end{equation}
Since the particle resides at the origin of $\bar{T}$, $\bar{x}$ is zero which simplifies Eq. \eqref{spaceTimeElementReduced} to
\begin{equation}
c\bar{t} = \sqrt{c^2 t^2 - x^2}.
\end{equation}
Furthermore, the velocity of the particle in $T$ is given by the relative speed $v$ between the two reference frames.
This provides us with a relation between $x$ and $t$: $x/t=v$.
Using this, we can write
\begin{equation}
c\bar{t} = \sqrt{c^2 t^2 - x^2} = \sqrt{1 - \frac{x^2}{c^2t^2}} ct = \sqrt{1 - \frac{v^2}{c^2}}  ct,
\end{equation}
and
\begin{equation} \label{timeDilation}
\bar{t} = \sqrt{1 - \frac{v^2}{c^2}} t.
\end{equation}
This special case of the transformation has to be consistent with our more general transformation given by Eq. \eqref{transformationBasedOnb}.
Thus the following two equations must hold for our special case:
\begin{equation}
\bar{t} = \sqrt{1+b^2}  t + \frac{b}{c} x = \sqrt{1+b^2}  t + \frac{b}{c} v t,
\end{equation}
and
\begin{equation}
\bar{t} = \sqrt{1 - \frac{v^2}{c^2}} t.
\end{equation}
This results in the following condition for b:
\begin{equation}
\sqrt{1 - \frac{v^2}{c^2}} t = \sqrt{1+b^2} t + \frac{b}{c} v t.
\end{equation}
As this equation must hold for arbitrary time $t$, we can write
\begin{equation}
\sqrt{1 - \frac{v^2}{c^2}} = \sqrt{1+b^2} + \frac{b}{c} v. \label{find_b}
\end{equation}
Solving Eq. \eqref{find_b} for $b$, we find that
\begin{equation}
b = \frac{-v/c}{\sqrt{1 - \frac{v^2}{c^2}}},
\end{equation}
which can be verified as:
\begin{equation}
\sqrt{1+b^2} + \frac{b}{c} v
= \sqrt{1+\frac{v^2/c^2}{1 - \frac{v^2}{c^2}}} - \frac{v^2/c^2}{\sqrt{1 - \frac{v^2}{c^2}}}
= \frac{1}{\sqrt{1 - \frac{v^2}{c^2}}}  - \frac{v^2/c^2}{\sqrt{1 - \frac{v^2}{c^2}}}
=  \sqrt{1 - \frac{v^2}{c^2}}.
\end{equation}

\subsubsection{Lorentz transformation} \label{sectLorentzTransform}

In preparation to use our representation of $b$ into Eq. \eqref{transformationBasedOnb}, we first consider
\begin{equation}
\sqrt{1+b^2}
= \sqrt{1 + \frac{v^2/c^2}{1 - \frac{v^2}{c^2}}}
= \sqrt{\frac{1 - v^2/c^2 + v^2/c^2}{1 - \frac{v^2}{c^2}}}
= \frac{1}{\sqrt{1 - \frac{v^2}{c^2}}}.
\end{equation}
Using this, Eq. \eqref{transformationBasedOnb} turns into
\begin{equation} \label{lorentz}
\left(\begin{array}{c}
\bar{t}
\\
\bar{x}
\end{array} \right)
=
\begin{pmatrix}
\frac{1}{\sqrt{1 - \frac{v^2}{c^2}}} & \frac{-v/c^2}{\sqrt{1 - \frac{v^2}{c^2}}}
\\
\frac{-v}{\sqrt{1 - \frac{v^2}{c^2}}}  & \frac{1}{\sqrt{1 - \frac{v^2}{c^2}}}
\end{pmatrix}
\left(\begin{array}{c}
t
\\
x
\end{array} \right).
\end{equation}
Eq. \eqref{lorentz} is called the Lorentz transformation.

\subsubsection{Interpretation and generalization} \label{sectionGeneralizationLorentz}

This result is valid with rigor for the IRF $\bar{T}$ in which the particle is at rest at the IRF's origin.
Even if the particle is accelerated along the $x$ axis, for small time intervals there always exists such an IRF.
Those IRFs are called the particle's \textit{momentary rest} IRFs.

If we look at Eq. \eqref{lorentz} in its full 4 dimensional form
\begin{equation}
\left(\begin{array}{c}
              \bar{t}
              \\
              \bar{x}
              \\
              \bar{y}
              \\
              \bar{z}
\end{array} \right)
=
\begin{pmatrix}
    \frac{1}{\sqrt{1 - \frac{v^2}{c^2}}} & \frac{-v/c^2}{\sqrt{1 - \frac{v^2}{c^2}}} & 0 & 0
    \\
    \frac{-v}{\sqrt{1 - \frac{v^2}{c^2}}}  & \frac{1}{\sqrt{1 - \frac{v^2}{c^2}}} & 0 & 0
    \\
    0 & 0 & 1 & 0
    \\
    0 & 0 & 0 & 1
\end{pmatrix}
\left(\begin{array}{c}
          t
          \\
          x
          \\
          y
          \\
          z
\end{array} \right),
\end{equation}
we can then multiply by spatial rotation matrices
\begin{equation} \label{lorentzWithRotation}
    \left(\begin{array}{c}
              \bar{t}
              \\
              \bar{x}
              \\
              \bar{y}
              \\
              \bar{z}
    \end{array} \right)
    =
    \begin{pmatrix}
        \frac{1}{\sqrt{1 - \frac{v^2}{c^2}}} & \frac{-v/c^2}{\sqrt{1 - \frac{v^2}{c^2}}} & 0 & 0
        \\
        \frac{-v}{\sqrt{1 - \frac{v^2}{c^2}}}  & \frac{1}{\sqrt{1 - \frac{v^2}{c^2}}} & 0 & 0
        \\
        0 & 0 & 1 & 0
        \\
        0 & 0 & 0 & 1
    \end{pmatrix}
    \begin{pmatrix}
        1 & 0 & 0 & 0
        \\
        0 & r_{11} & r_{12} & r_{13}
        \\
        0 & r_{21} & r_{22} & r_{23}
        \\
        0 & r_{31} & r_{32} & r_{33}
    \end{pmatrix}
    \left(\begin{array}{c}
              t
              \\
              x
              \\
              y
              \\
              z
    \end{array} \right)
\end{equation}
(where $r_{ij}$ with $ i,j \in \{1,2,3\}$ denote the components of the rotation matrix) and note that the product still fulfills Eq. \eqref{spaceTimeElement}.
With this in mind, we can claim that our derivation is valid for the momentary rest IRF of a particle moving in an arbitrary direction relative to $T$.

With less rigor, we can claim that Eq. \eqref{lorentz} holds when $\bar{T}$ is not the particle's rest IRF, but another IRF from which the particle is observed.
Even in this case, Eq. \eqref{lorentz} fulfills the central proposition in Eq. \eqref{spaceTimeElement} and thus is a good candidate for being the correct transformation.

\paragraph{Translations in time and space}
By requiring $T$ and $\bar{T}$ to be on top of each other at time $t=\bar{t}=0$, we excluded time and spatial translations.
This restriction is not relevant if we are considering coordinate differences where translations cancel.
Specifically, for coordinate differences, the transformation
\begin{equation} \label{lorentzWithRotationForDifferences}
\left(\begin{array}{c}
          \Delta \bar{t}
          \\
          \Delta \bar{x}
          \\
          \Delta \bar{y}
          \\
          \Delta \bar{z}
\end{array} \right)
=
\begin{pmatrix}
    \frac{1}{\sqrt{1 - \frac{v^2}{c^2}}} & \frac{-v/c^2}{\sqrt{1 - \frac{v^2}{c^2}}} & 0 & 0
    \\
    \frac{-v}{\sqrt{1 - \frac{v^2}{c^2}}}  & \frac{1}{\sqrt{1 - \frac{v^2}{c^2}}} & 0 & 0
    \\
    0 & 0 & 1 & 0
    \\
    0 & 0 & 0 & 1
\end{pmatrix}
\begin{pmatrix}
    1 & 0 & 0 & 0
    \\
    0 & r_{11} & r_{12} & r_{13}
    \\
    0 & r_{21} & r_{22} & r_{23}
    \\
    0 & r_{31} & r_{32} & r_{33}
\end{pmatrix}
\left(\begin{array}{c}
          \Delta t
          \\
          \Delta x
          \\
          \Delta y
          \\
          \Delta z
\end{array} \right)
\end{equation}
is general enough for the purpose of this paper.

For the case of absolute coordinates with Eq. \eqref{lorentzWithRotation}, we will always be able to reach an IRF which is at relative rest to another IRF.
From Eq. \eqref{noVelocityTransform}, we know that relativistic effects occur only when IRFs are moving relative to each other.

\section{Concepts and notations}

\subsection{Proper time} \label{sectionProperTime}

Let $\bar{T}$, with coordinates $\bar{t}$, $\bar{x}_1$, $\bar{x}_2$, $\bar{x}_3$, be a particle's momentary rest IRF, and let $T$, with coordinates $t$, $x_1$, $x_2$, $x_3$, be an IRF from which we observe the particle.
Then
\begin{equation}
    c^2\dd \bar{t}^2 - \dd \bar{x}^2 = c^2\dd t^2 - \dd x^2
\end{equation}
with $\bar{x} := (\bar{x}_1,\bar{x}_2,\bar{x}_3)$ and $x := (x_1,x_2,x_3)$ holds.

Because $\dd\bar{x}=0$ in the particle's momentary rest IRF, we are left with
\begin{equation} \label{defProperTime}
c \dd \bar{t} = \sqrt{c^2 \dd t^2 - \dd x^2} = c \sqrt{1- \frac{v^2}{c^2}} \dd t \iff \dd \bar{t} = \sqrt{1- \frac{v^2}{c^2}} \dd t
\end{equation}
where $v=v(t)$ is the particle's momentary velocity in the observer IRF.

According to its definition, $\dd \bar{t} = \sqrt{1-\frac{v^2}{c^2}} \dd t$ is the time interval in the particle's momentary rest IRF that corresponds to the time interval $\dd t$ in the observer IRF.
As a result, any observer can calculate how long a time interval $\dd t$ in their IRF will be in the particle's momentary rest IRF using $\sqrt{1-\frac{v^2}{c^2}} \dd t$.
Thus, all observers in IRFs will agree upon the time interval $\dd \bar{t}$ in the particle's momentary rest IRF.
Through this formula, $\dd \bar{t}$ is the same in any IRF. By convention, time intervals in the particle's momentary rest IRF are named $\dd\tau$ and are called \textit{proper time}:
\begin{equation}
    \dd\tau = \sqrt{1-\frac{v^2}{c^2}}  \dd t.
\end{equation}

A more formal, and even simpler, way to see that $\dd\tau$ is the same in every IRF is as follows.
The value of
\begin{equation}
    c^2\dd t^2 -\dd x^2 = c^2\dd t^2 -\dd {x_1}^2-\dd {x_2}^2 -\dd {x_3}^2
\end{equation}
is the same in all IRFs.
The same is true for its square root:
\begin{equation}
    \sqrt{c^2 \dd t^2 - \dd x^2} = \sqrt{1-\frac{v^2}{c^2}} \; \dd t = \dd\tau.
\end{equation}

\subsection{4-velocity} \label{section4Velocity}
The invariance of proper time $\dd \tau$ allows us to define a new object which under Lorentz transformations transforms like $(c  \dd t, \dd x_1,\dd x_2, \dd x_3)$:
\begin{equation}\label{4velocityu}
    u := \frac{1}{\dd \tau}
    \left(\begin{array}{c}
              c \dd t\\
              \dd x_1\\
              \dd x_2\\
              \dd x_3\\
    \end{array} \right)
    = \frac{1}{\sqrt{1-\frac{v^2}{c^2}}}
    \left(\begin{array}{c}
              c \\
              v_1\\
              v_2\\
              v_3\\
    \end{array} \right).
\end{equation}
The velocity $u$ is called the particle's ``4-velocity.''

\subsection{Notations and 4-vectors} \label{sectionNotations}
In Eq. \eqref{lorentzWithRotationForDifferences} we gave the Lorentz transformation for differences of the coordinates $(t,x,y,z)$.
In this section we take the coordinates to be of the form $(c t,x_1,x_2,x_3)$.
This will change the Lorentz transformation slightly: it will require reverting what we did in Eq. \eqref{shiftCFromVectorToMatrix} to Eq. \eqref{transformationBasedOnb}.
For the rest of this paper we will be interested in the Lorentz transformation based on $(c t,x_1,x_2,x_3)$.

From Eq. \eqref{lorentzWithRotationForDifferences}, we know that Lorentz transformations can be written as $4 \times 4$ matrices.
In the following we denote such matrices by $\Lambda$.

For any Lorentz transformation $\Lambda$ of $\Delta X := (c \Delta t, \Delta x_1, \Delta x_2, \Delta x_3)$ the following equation must hold
\begin{equation} \label{generalConditionForLorentzTransformations}
\Lambda^\mathsf{T} g \Lambda = g \;\; \text{where} \;\;
g :=
\begin{pmatrix}
    1 & 0 & 0 & 0
    \\
    0 & -1 & 0 & 0
    \\
    0 & 0 & -1 & 0
    \\
    0 & 0 & 0 & -1
\end{pmatrix}.
\end{equation}
In special relativity, $g$ is called the metric tensor or simply the metric.
Eq. \eqref{generalConditionForLorentzTransformations} is a direct consequence of Eq. \eqref{invarianceSpaceTimeInterv}; it is the formal definition of a Lorentz transformation, i.e. any $4 \times 4$ matrix that fulfills this equation is a Lorentz transformation of $\Delta X$.
With these definitions, the transformation of $\Delta X$ by $\Lambda$ is given by matrix-vector multiplication $\Lambda \left(\Delta X \right)$.
Up to now we know two 4-component objects that transform like $\Delta X$.
Those are $\Delta X$ itself and the 4-velocity $u$ defined in Eq. \eqref{4velocityu}.
Any 4-component object that transforms like $\Delta X$ is called a ``4-vector.''

\subsection{Invariance of the 4-vector product} \label{invarianceOf4VectorProduct}
The product $V^\mathsf{T} g W$ of two 4-vectors $V,W$ is invariant under Lorentz transformations, i.e. is the same in any IRF:
\begin{equation}
    (\Lambda V)^\mathsf{T} g (\Lambda W) = V^\mathsf{T} \Lambda^\mathsf{T} g \Lambda W = V^\mathsf{T} g W.
\end{equation}

\subsection{Proper time revisited}
With the concept of the 4-vector product we can write
\begin{align}
    & \dd X^\mathsf{T} g \; \dd X = c^2 \dd t^2 - \dd x^2 \\
    \iff & \sqrt{\dd X^\mathsf{T} g \; \dd X} = c \; \sqrt{1-\frac{v^2}{c^2}} \; \dd t = c \; \dd \tau
\end{align}
This is another proof of the invariance of $\dd \tau$ under Lorentz transformations.

\section{On the Lagrange formulation of particle dynamics} \label{particleLagrangian}

The main result of this section will be the derivation of the transformation law of the electromagnetic potentials $A$ and $\phi$, as given in Ref.~\cite{WagnerGuthrieClassicalField}, from the invariance of the Lorentz force.
Another important result will be the derivation of the famous mass-energy equivalence formula $E=mc^2$.

Quite a few preparations in the field of classical non-relativistic particle physics will be needed.
These preparations are done in the section \ref{subsectionPreparations}.

\subsection{On the Lagrange formulation of classical non relativistic particle dynamics}\label{subsectionPreparations}

\subsubsection{Energy conservation} \label{energyConservation}
The Lagrangian formalism for particle physics, as described in Ref.~\cite{WagnerGuthrie}, allows us to derive energy conservation from analyzing how the action $S$ behaves under infinitesimal time translations. 
By doing so, we will find a definition for the energy of any physical system that is described by a particle Lagrangian. \\

To begin, we consider the action

\begin{equation}
S = \int_{t_1}^{t_2} L\left(q(t), \dot{q}(t), t\right) \dd t,
\end{equation}
where $L(q(t), \dot{q}(t), t)$ is the particle's Lagrange function, $q$ is the particle's position coordinates, $\dot{q}$ is the particle's velocity, and $t$ is time. 
$t_1$ and $t_2$ are fixed but arbitrary endpoints of a time interval of which we calculate the particle's action $S$ ($S$ can be considered as a function of $t_1$ and $t_2$: $S = S(t_1,t_2)$).

We now ask by what amount $S$ changes if time is changed from $t$ to $t + \delta t$, with $\delta t$ being a small time interval.
The resulting change $\delta S$ of $S$ is given by
\begin{equation} \label{defDelS}
\delta S = \int_{t_1 + \delta t}^{t_2 + \delta t} L(q(t), \dot{q}(t), t) \dd t 
- \int_{t_1}^{t_2} L(q(t), \dot{q}(t), t) \dd t.
\end{equation}
Next, we use the substitution rule which is given by
\begin{equation}
\int_{\varphi(t_1)}^{\varphi(t_2)} f(x) \dd x = \int_{t_1}^{t_2} f(\varphi(t)) \; \dot{\varphi}(t) \dd t.
\end{equation}
For $\varphi (t) = t + \delta t$ the rule reads
\begin{equation}
\int_{t_1 + \delta t}^{t_2 + \delta t} f(t) \dd t = \int_{t_1}^{t_2} f(t + \delta t) \dd t.
\end{equation}
Because $\delta t$ is small, we can write
\begin{equation}
\int_{t_1 + \delta t}^{t_2 + \delta t} f(t) \dd t 
= \int_{t_1}^{t_2} \left(f(t) + \frac{\dd f}{\dd t} \delta t \right) \dd t.
\end{equation} 
Applying this to Eq. \eqref{defDelS}, we arrive at
\begin{equation}
\delta S = \int_{t_1}^{t_2} \left(L + \frac{\dd L}{\dd t} \delta t \right) \dd t - \int_{t_1}^{t_2} L \dd t
= \delta t \int_{t_1}^{t_2} \frac{\dd L}{\dd t} \dd t.
\end{equation}

We will now assume that the particle's trajectory $q(t)$ is its physical trajectory which fulfills the Euler-Lagrange equations
\begin{equation} \label{ParticleEulLagEqu}
\frac{\dd}{\dd t} \frac{\partial L}{\partial \dot{q}} - \frac{\partial L}{\partial q} = 0.
\end{equation}
To make use of this equation we reformulate $\delta S$ as follows
\begin{equation}
\delta S = \delta t \int_{t_1}^{t_2} \frac{\dd L}{\dd t} \dd t
= \delta t \int_{t_1}^{t_2} \left(
\frac{\partial L}{\partial q} \dot{q} + \frac{\partial L}{\partial \dot{q}} \ddot{q} + \frac{\partial L}{\partial t} 
\right) \dd t.
\end{equation}
Integration by parts of the second term leads to
\begin{equation}
\delta S = \delta t \int_{t_1}^{t_2} \left(
\frac{\partial L}{\partial q} \dot{q} -\left( \frac{\dd }{\dd t}\frac{\partial L}{\partial \dot{q}} \right) \dot{q} 
 + \frac{\partial L}{\partial t}
\right) \dd t
+ \delta t \left[ \frac{\partial L}{\partial \dot{q}} \dot{q} \right]_{t_1}^{t_2}
\end{equation}
With Eq. \eqref{ParticleEulLagEqu}, we are left with
\begin{equation}
\delta S = \delta t \int_{t_1}^{t_2} \frac{\partial L}{\partial t} \dd t
+ \delta t \left[ \frac{\partial L}{\partial \dot{q}} \dot{q} \right]_{t_1}^{t_2}
= \delta t \int_{t_1}^{t_2} \frac{\partial L}{\partial t} \dd t
+ \delta t \int_{t_1}^{t_2} \frac{\dd}{\dd t} \left( \frac{\partial L}{\partial \dot{q}} \dot{q} \right) \dd t.
\end{equation}

We have now found two expressions for $\delta S$:
\begin{equation}
\delta S = \delta t \int_{t_1}^{t_2} \frac{\dd L}{\dd t} \dd t 
\;\; \text{and} \;\;
\delta S = \delta t \int_{t_1}^{t_2} \frac{\partial L}{\partial t} \dd t
+ \delta t \int_{t_1}^{t_2} \frac{\dd}{\dd t} \left( \frac{\partial L}{\partial \dot{q}} \dot{q} \right) \dd t.
\end{equation}
Setting these to equal gives
\begin{align}
\int_{t_1}^{t_2} \frac{\dd L}{\dd t} \dd t
&=  \int_{t_1}^{t_2} \frac{\partial L}{\partial t} \dd t
+ \int_{t_1}^{t_2} \frac{\dd}{\dd t} \left( \frac{\partial L}{\partial \dot{q}} \dot{q} \right) \dd t \\
\iff
- \int_{t_1}^{t_2} \frac{\partial L}{\partial t} \dd t  
&=  \int_{t_1}^{t_2} \left(
\frac{\dd}{\dd t} \left( \frac{\partial L}{\partial \dot{q}} \dot{q} \right) - \frac{\dd L}{\dd t} \right) \dd t \\
\iff
- \int_{t_1}^{t_2} \frac{\partial L}{\partial t} \dd t  
&=  \int_{t_1}^{t_2} \frac{\dd}{\dd t} \left(\frac{\partial L}{\partial \dot{q}} \dot{q}  - L \right) \dd t.
\end{align}
For the special case where $\partial L / \partial t = 0 $, this leads to
\begin{equation}
0 = \int_{t_1}^{t_2} \frac{\dd}{\dd t} \left(\frac{\partial L}{\partial \dot{q}} \dot{q}  - L \right) \dd t.
\end{equation}
Since $t_1$ and $t_2$ are arbitrary, this equation can only be true if
\begin{equation} \label{legendre}
0 = \frac{\dd}{\dd t} \left(\frac{\partial L}{\partial \dot{q}} \dot{q}  - L \right) 
\implies 
E := \frac{\partial L}{\partial \dot{q}} \dot{q}  - L = \mathrm{constant}.
\end{equation}

\paragraph{Interpretation}

These derivations result in the following collection of interpretations:

\begin{itemize}
\item $E$ is called energy of the particle. $E$ is defined for any particle that can be described by the Lagrangian formalism for particle physics. It is constant in time when $\partial L / \partial t = 0$, i.e. when the particle's Lagrangian does not explicitly depend on time but depends on time through the particle's coordinates $q(t)$ only.

\item 
$E$ was derived from analyzing how $S$ behaves under infinitesimal time translation. 
Energy conservation can then be considered as a consequence of the behavior of a particle's action $S$ under infinitesimal time translations.

\item
There are more transformations, e.g. spatial translations and rotations, which also lead to conserved quantities. 
Those are easier to derive because for them $\delta S = 0$.
They can be found in any classical mechanics text book, a common example of which can be found in Goldstein~\cite{GoldsteinConservation}.

\item
For the standard classical mechanics Lagrangian
\begin{equation}
L = T-V = \frac{1}{2} m v^2 - V ,
\end{equation}
the energy calculates to
\begin{equation}
E = \frac{\partial}{\partial v}\left[ \left(\frac{1}{2} m v^2 - V \right) v\right] - \left(\frac{1}{2} m v^2 - V\right) =  \frac{1}{2} m v^2 + V.
\end{equation}

\end{itemize}

\subsubsection{The Lorentz force and its Lagrangian} \label{sectionLorentzForceLagrangian}

The Lorentz force on a particle with charge $e$ in Cartesian coordinates is given by
\begin{equation} \label{lorentzForceLaw}
F_L = e E + e v \times B,
\end{equation}
where $v$ is the particle's velocity and $E$ and $B$ are the electric and magnetic fields at the particle's coordinates, respectively.
The Lagrangian that corresponds to the Lorentz force is given by
\begin{equation} \label{lorentzForceLagrangian}
L_L = - e (\phi - A \cdot v),
\end{equation}
where $\phi$ and $A$ are respectively the electric and magnetic potentials as discussed in Ref.~\cite{WagnerGuthrieClassicalField}.

$L_L$ is the part of the Lagrangian of a charged particle in an electromagnetic field that represents the particle's interaction with the electromagnetic field.
In classical mechanics the complete Lagrangian is
\begin{equation}
    L = \frac{1}{2}mv^2 + L_L.
\end{equation}
To show that $L_L$ reproduces the Lorentz force $F_L$ we prove
\begin{equation}
    F_L = -\left(\frac{\dd}{\dd t} \frac{\partial L_L}{\partial v} - \frac{\partial L_L}{\partial x} \right).
\end{equation}
We start with
\begin{align}
    \frac{\partial L_L}{\partial v_i} &= e A_i \\
    \implies \frac{\dd}{\dd t} \frac{\partial L_L}{\partial v_i} &= e \left(\frac{\partial A_i}{\partial t} + \frac{\partial A_i}{\partial x_j} v_j\right).
\end{align}
with $ i,j \in \{1,2,3\}$ and implicit sum over duplicate indices.
To understand the term $\frac{\partial A_i}{\partial x_j} v_j$, we consider that during some small time interval $\dd t$ the particle's coordinates change by $\dd x_j = v_j \dd t$.
Hence, $A_i$'s change resulting from the change $\dd x_j$ is given by $\frac{\partial A_i}{\partial x_j} \dd x_j = \frac{\partial A_i}{\partial x_j} v_j \dd t$.
The coordinate-wise contribution of $A_i$ to the total time derivative of $\frac{\partial L_L}{\partial v_i}$ is $\frac{\partial A_i}{\partial x_j} v_j$.

We continue with $\frac{\partial L_L}{\partial x_i}$:

\begin{equation}
    \frac{\partial L_L}{\partial x_i} = -e \frac{\partial \phi}{\partial x_i} + e \frac{\partial A_j}{\partial x_i} v_j.
\end{equation}
Thus,
\begin{align}
    \frac{\dd}{\dd t} \frac{\partial L_L}{\partial v_i} - \frac{\partial L_L}{\partial x_i}
    & = e \left(\frac{\partial A_i}{\partial t} + \frac{\partial A_i}{\partial x_j} v_j\right)
    - \left( -e \frac{\partial \phi}{\partial x_i} + e \frac{\partial A_j}{\partial x_i} v_j \right) \nonumber \\
    & = - e \left( -\frac{\partial \phi}{\partial x_i} - \frac{\partial A_i}{\partial t} \right)
    - e \left( v_j \frac{\partial A_j}{\partial x_i}  - v_j \frac{\partial A_i}{\partial x_j}  \right).
\end{align}
We next use the relations $B = \nabla \times A$ and $E = - \nabla \phi - \frac{\partial A}{\partial t}$, which were introduced in Ref.~\cite{WagnerGuthrieClassicalField}.

As an intermediate step, we consider
\begin{align}
[v \times B]_i &= [v \times (\nabla \times A)]_i \nonumber \\
&=\epsilon_{ijk} v_j [\nabla \times A]_k \nonumber \\
&=\epsilon_{ijk} v_j \epsilon_{kln} \frac{\partial A_n}{\partial x_l} \nonumber \\
&=\epsilon_{kij} \epsilon_{kln} v_j  \frac{\partial A_n}{\partial x_l} \nonumber \\
&=(\delta_{il} \delta_{jn} - \delta_{in} \delta_{jl} ) v_j  \frac{\partial A_n}{\partial x_l} \nonumber \\
&=v_j \frac{\partial A_j}{\partial x_i} - v_j \frac{\partial A_i}{\partial x_j} \nonumber,
\end{align}
where $\epsilon_{ijk}$ is the Levi-Civita symbol, $\delta_{ij}$ is the Kronecker delta, and where we made use of the rule
$\epsilon_{ijk} \epsilon_{ilm} = \delta_{jl}\delta_{km} - \delta_{jm}\delta_{kl}$.

With that, we arrive at
\begin{equation}
    \frac{\dd}{\dd t} \frac{\partial L_L}{\partial v_i} - \frac{\partial L_L}{\partial x_i}
    = - e E_i - e [v \times B]_i = -{F_L}_i,
\end{equation}
which is the result we wanted to prove.

\subsection{On the Lagrange formulation of relativistic particle dynamics} \label{sectionOnRelativisticParticles}

\subsubsection{Invariance of the relativistic particle Euler-Lagrange equations} \label{relativisticParticleEulerLagrangeInvariance}
\paragraph{Preliminaries and notations}
The invariance of the Euler-Lagrange equations under wide ranges of transformations has always been our crucial argument to accept the Lagrangian formalism.
We already used this argument to introduce and rectify the Lagrangian formalism for classical mechanics~\cite{WagnerGuthrie} and classical field theory~\cite{WagnerGuthrieClassicalField}.
In this paragraph, we are going to discuss under which conditions the particle Euler-Lagrange equations are invariant under transformations which include time.
Of course, the transformations we are especially interested in are Lorentz transformations.

Let $(t,x); \; x:=(x_1,x_2,x_3)$ and $(\bar{t}, \bar{x}); \; \bar{x}:=(\bar{x}_1,\bar{x}_2,\bar{x}_3)$ be time and spatial coordinates.
An invertible and differentiable transformation between the coordinates is then given by
\begin{equation} \label{particleTransformWithTime}
\left(\begin{array}{c}
          t \\
          x
\end{array} \right)
=
\left(\begin{array}{c}
          \varphi(\bar{t}, \bar{x}) \\
          f(\bar{t}, \bar{x})
\end{array} \right).
\end{equation}
Our aim is to show that the principle of stationary action for a particle action of the form
\begin{equation}
    S = \int_{t_1}^{t_2} L \left( x(t), \frac{\dd x}{\dd t}(t), t \right) \dd t
\end{equation}
leads to Euler-Lagrange equations, which are form invariant under Eq. \eqref{particleTransformWithTime}.

For a given particle trajectory $\bar{x}(\bar{t})$,  the functions $\varphi$ and $f$ can be written as
\begin{align}
    \varphi(\bar{t}) & := \varphi(\bar{t}, \bar{x}(\bar{t})) \label{timeTransformTOnly} \\
    f(\bar{t}) & :=  f(\bar{t}, \bar{x}(\bar{t})).
\end{align}

\paragraph{Definition of the transformation of the Lagrangian}
We define the transformation of the Lagrangian under $\varphi$ and $f$ by
\begin{equation} \label{definitionParticleLagrangianTransformationWithTime}
\bar{L} \left(\bar{x}(\bar{t}), \frac{\dd \bar{x}}{\dd \bar{t}}, \bar{t} \right) := L \left(f, \frac{\dd f}{\dd \varphi},\varphi \right) \; \frac{\dd \varphi}{\dd \bar{t}} .
\end{equation}
The particle's action can then be written in the following two ways:
\begin{align}
    S & = \int_{\bar{t}_1}^{\bar{t}_2} \bar{L} \left(\bar{x}(\bar{t}), \frac{\dd \bar{x}}{\dd \bar{t}}, \bar{t} \right) \dd \bar{t}, \label{relativisticParticleActionTransformed1} \\
    \iff S & = \int_{\bar{t}_1}^{\bar{t}_2} L \left(f, \frac{\dd f}{\dd \varphi},\varphi \right) \; \frac{\dd \varphi}{\dd \bar{t}} \; \dd \bar{t}. \label{relativisticParticleActionTransformed2}
\end{align}
For Eq. \eqref{relativisticParticleActionTransformed1}, the Euler-Lagrange equations
\begin{equation} \label{relativisticParticleEulerLagrangeBarred}
\frac{\dd}{\dd \bar{t}} \frac{\partial \bar{L}}{\partial \left(\frac{\dd \bar{x}}{\dd \bar{t}}\right)} - \frac{\partial \bar{L}}{\partial \bar{x}} = 0
\end{equation}
follow from applying the principle of stationary action as discussed in Ref.~\cite{WagnerGuthrie}.

\paragraph{Proof that $\frac{\dd}{\dd t} \frac{\partial L}{\partial \left(\frac{\dd x}{\dd t}\right)} - \frac{\partial L}{\partial x} = 0$ holds} \label{sectionProofEulerLagrangeRelPartcle1}
Before we apply the principle of stationary action to the particle's action in Eq. \eqref{relativisticParticleActionTransformed2}, we first use the substitution rule
\begin{equation}
    \int_{\varphi(\bar{t}_1)}^{\varphi(\bar{t}_2)} L \dd \varphi
    = \int_{\bar{t}_1}^{\bar{t}_2} L  \; \frac{\dd \varphi}{\dd \bar{t}} \; \dd \bar{t}.
\end{equation}
This requires that every dependence on $\bar{t}$ in $L$ must be replaced by $\varphi$.
Only then can $\varphi$ become the new integration variable.
As $L= L \left(f, \frac{\dd f}{\dd \varphi},\varphi \right)$, this means that $f(\bar{t})$ needs to be written as a function of $\varphi$.
The only way this could be possible  in general  is if $\varphi$ can be inverted:
\begin{equation}
    \bar{t} = \varphi^{-1} (\varphi),
\end{equation}
where $\varphi^{-1}$ denotes a function while $\varphi$ is just a variable, namely the integration variable.
Unfortunately, we only assumed Eq. \eqref{particleTransformWithTime} to be invertible.
From Eq. \eqref{particleTransformWithTime} being invertible, it does not follow that $\varphi(\bar{t})$ is invertible, i.e. that the function $\varphi^{-1}$ exists.

\paragraph{Important examples of invertible $\varphi(\bar{t})$} \label{sectionExampleOfInvertiblePhi}
The aim of this paragraph is to motivate that it is rectified to claim $\varphi(\bar{t})$ to be invertible:
The important examples for transformations of the form \eqref{particleTransformWithTime} are Lorentz transformations.
So we explore what $\varphi(\bar{t})$ looks like for Lorentz transformations:

Let $T$ and $\bar{T}$ be IRFs with time and spatial coordinates $(t,x)$ and $(\bar{t},\bar{x})$ respectively.
Inspired by section \ref{sectionGeneralizationLorentz}, we consider the following transformation of differences of $(\bar{t},\bar{x})$:
\begin{align}
& \begin{pmatrix}
    \frac{1}{\sqrt{1 - \frac{v^2}{c^2}}} & \frac{-v/c^2}{\sqrt{1 - \frac{v^2}{c^2}}} & 0 & 0
    \\
    \frac{-v}{\sqrt{1 - \frac{v^2}{c^2}}}  & \frac{1}{\sqrt{1 - \frac{v^2}{c^2}}} & 0 & 0
    \\
    0 & 0 & 1 & 0
    \\
    0 & 0 & 0 & 1
\end{pmatrix}
\begin{pmatrix}
    1 & 0 & 0 & 0
    \\
    0 & r_{11} & r_{12} & r_{13}
    \\
    0 & r_{21} & r_{22} & r_{23}
    \\
    0 & r_{31} & r_{32} & r_{33}
\end{pmatrix}
\left(\begin{array}{c}
          \Delta \bar{t}
          \\
          \Delta \bar{x}_1
          \\
          \Delta \bar{x}_2
          \\
          \Delta \bar{x}_3
\end{array} \right) \nonumber \\
= & \begin{pmatrix}
        \frac{1}{\sqrt{1 - \frac{v^2}{c^2}}} & \frac{-v/c^2}{\sqrt{1 - \frac{v^2}{c^2}}} r_{11} & \frac{-v/c^2}{\sqrt{1 - \frac{v^2}{c^2}}} r_{12} & \frac{-v/c^2}{\sqrt{1 - \frac{v^2}{c^2}}} r_{13}
        \\
        \frac{-v}{\sqrt{1 - \frac{v^2}{c^2}}}  & \frac{1}{\sqrt{1 - \frac{v^2}{c^2}}} r_{11} & \frac{1}{\sqrt{1 - \frac{v^2}{c^2}}} r_{12} & \frac{1}{\sqrt{1 - \frac{v^2}{c^2}}} r_{13}
        \\
        0 & r_{21} & r_{22} & r_{23}
        \\
        0 & r_{31} & r_{32} & r_{33}
\end{pmatrix}
\left(\begin{array}{c}
          \Delta \bar{t}
          \\
          \Delta \bar{x}_1
          \\
          \Delta \bar{x}_2
          \\
          \Delta \bar{x}_3
\end{array} \right), \label{explicitLorentzTransformWithRotation}
\end{align}
where $r_{ij}$ are the elements of a $3 \times 3$ rotation matrix embedded in the lower right of a $4 \times 4$ matrix and $v$ is defined as the absolute value of the relative velocity between $T$ and $\bar{T}$.
The rotation is chosen in such a way that it aligns the $x_1$ axis of $\bar{T}$ with the vector of the relative velocity between $T$ and $\bar{T}$.

With this transformation, we reach a system which is relative to $T$ at rest.
From Eq. \eqref{noVelocityTransform}, we know that the transformation from this system to $T$ does not imply any transformation of time.

Thus, the complete transformation of time between $\bar{T}$ and $T$ can be calculated from Eq. \eqref{explicitLorentzTransformWithRotation} and is given by
\begin{equation}
    \Delta t = \frac{1}{\sqrt{1 - \frac{v^2}{c^2}}} \left(\Delta \bar{t} - \frac{v \;\; \sum_{i=1}^3 r_{1i} \;\Delta\bar{x}_i }{c^2}\right).
\end{equation}
As $r_{ij}$ is chosen in such a way that $\sum_{i=1}^3 r_{1i}\Delta\bar{x}_i$ is the complete spatial distance $T$ moves relative to $\bar{T}$ in time $\Delta \bar{t}$, the relation $v = \frac{\sum_{i=1}^3 r_{1i} \Delta\bar{x}_i}{\Delta \bar{t}}$ holds.
With this, we arrive at
\begin{equation}
    \Delta t = \frac{1}{\sqrt{1 - \frac{v^2}{c^2}}} \left(\Delta \bar{t} - \frac{v^2 \; \Delta \bar{t} }{c^2}\right)
    = \sqrt{1 - \frac{v^2}{c^2}} \;\; \Delta \bar{t}.
\end{equation}

If we write this formula for $\bar{T}$ as the momentary rest IRFs of the particle, which in general we allow to be accelerating, $v$ will change with $\bar{t}$ and the differences will have to be turned into differentials:
\begin{equation}
    \dd t = \sqrt{1 - \frac{v(\bar{t})^2}{c^2}} \;\; \dd \bar{t}.
\end{equation}
Next, we rewrite this using the $t = \varphi(\bar{t})$ notation from Eqs. \eqref{particleTransformWithTime} and \eqref{timeTransformTOnly}:
\begin{equation} \label{derivationOfPhiWithRespectToTBar}
\dd \varphi = \sqrt{1 - \frac{v(\bar{t})^2}{c^2}} \;\; \dd \bar{t}
\iff \frac{\dd \varphi}{\dd \bar{t}} = \sqrt{1 - \frac{v(\bar{t})^2}{c^2}}.
\end{equation}
This is the desired result: For a particle that moves with velocity $v<c$, the derivative of $\varphi$ is always positive.
Thus, the function $\varphi^{-1}$ exists.

To summarize: For Lorentz transformations to a particle's momentary rest IRFs, we proved that the function $\varphi(\bar{t})$ in Eq. \eqref{timeTransformTOnly} is invertible.
This example provides evidence that it is reasonable to continue the former section assuming that $\varphi^{-1}$ exists.

\paragraph{Proof that $\frac{\dd}{\dd t} \frac{\partial L}{\partial \left(\frac{\dd x}{\dd t}\right)} - \frac{\partial L}{\partial x} = 0$ holds, continued} \label{sectionProofEulerLagrangeRelPartcle2}
Since we can now assume that $\varphi^{-1} = \varphi^{-1}(\varphi)$ exists, and that the dependence of $f$ on $\bar{t}$ can be replaced by $\bar{t} = \varphi^{-1}(\varphi)$, we know that the substitution rule makes sense and continue with
\begin{equation} \label{relativisticParticleActionAfterSubstition}
S= \int_{\varphi(\bar{t}_1)}^{\varphi(\bar{t}_2)} L\left( f, \frac{\dd f}{\dd \varphi}, \varphi \right)  \dd \varphi.
\end{equation}
At this point we need to recall from Ref.~\cite{WagnerGuthrie} that for the proof of Eq. \eqref{relativisticParticleEulerLagrangeBarred} we needed to consider an arbitrary variation $\delta \bar{x}(\bar{t})$ of the particle's trajectory in the barred coordinates, which vanished at the end points $\bar{t}_1$ and $\bar{t}_2$ of the action integral: $\delta \bar{x}(\bar{t}_1) = \delta \bar{x}(\bar{t}_2) = 0$.
The same variation of the trajectory in the unbarred coordinates is given by
\begin{equation} \label{deltaFByDeltaX}
\delta x = \delta f = \frac{\partial f}{\partial \bar{x} }  \delta \bar{x}.
\end{equation}
As the variation of the trajectory vanishes at $\bar{t}_1$ and $\bar{t}_2$ in the barred coordinates, in the unbarred coordinates the variation $\delta f$ vanishes at $t_1=\varphi(\bar{t}_1)$ and $t_2=\varphi(\bar{t}_2)$: \footnote{The simplest way to picture these equations is to imagine a real trajectory in space, which is described from within two systems of coordinates.}
\begin{equation} \label{borderConditionForF}
0 = \delta f(\varphi(\bar{t}_1)) = \delta f(\varphi(\bar{t}_2)).
\end{equation}
Now we are ready to use $\delta f$ to calculate the condition for the action $S$ to become stationary:
The variation $\delta S$ of the action shown in Eq.  \eqref{relativisticParticleActionAfterSubstition} resulting from the variation $\delta f$ is given by
\begin{equation} \label{relativisticParticleActionAfterSubstitionVariation}
\delta S = \int_{\varphi(\bar{t}_1)}^{\varphi(\bar{t}_2)} \frac{\partial L}{\partial f} \delta f
+ \frac{\partial L}{\partial \left(\frac{\dd f}{\dd \varphi}\right)} \delta \left( \frac{\dd f}{\dd \varphi} \right)  \dd \varphi.
\end{equation}
\\ \\
\textbf{Note:} In Eq. \eqref{relativisticParticleActionTransformed2}, $\varphi$ was still a function $\varphi = \varphi(\bar{t}) = \varphi(\bar{t},\bar{x}(\bar{t}))$.
By the substitution rule, $\varphi$ became the integration variable, allowing us to write $\delta S$ without considering variations of $\varphi$.
With that,we are now able to derive the Euler-Lagrange equations in the well known manner used in Ref.~\cite{WagnerGuthrie}.
This makes it clear how crucial $\frac{\dd \varphi}{\dd \bar{t}}$ was in definition \eqref{definitionParticleLagrangianTransformationWithTime}.
Without $\frac{\dd \varphi}{\dd \bar{t}}$ we would be unable to derive invariant Euler-Lagrange equations and the Lagrangian formalism would essentially break down.
\\ \\
Integration by parts turns Eq. \eqref{relativisticParticleActionAfterSubstitionVariation} into
\begin{equation}
    \delta S = \int_{\varphi(\bar{t}_1)}^{\varphi(\bar{t}_2)} \frac{\partial L}{\partial f} \delta f
    - \frac{\dd}{\dd \varphi} \left(\frac{\partial L}{\partial \left(\frac{\dd f}{\dd \varphi}\right)}\right) \! \delta f   \dd \varphi
    + \left[ \frac{\partial L}{\partial \left(\frac{\dd f}{\dd \varphi}\right)} \; \delta f \right]_{\varphi(\bar{t}_1)}^{\varphi(\bar{t}_2)}
\end{equation}
Because of the boundary condition defined in Eq. \eqref{borderConditionForF}, the last summand vanishes and we are left with
\begin{equation}
    \delta S = \int_{\varphi(\bar{t}_1)}^{\varphi(\bar{t}_2)} \left(\frac{\partial L}{\partial f}
    - \frac{\dd}{\dd \varphi} \left(\frac{\partial L}{\partial \left(\frac{\dd f}{\dd \varphi}\right)}\right) \right) \delta f \dd \varphi .
\end{equation}
Also, because of the relation in Eq. \eqref{deltaFByDeltaX}, $\delta f$ is as arbitrary as $\delta \bar{x}$.
The condition for $S$ to be stationary, which is equivalent to $\delta S = 0$, is given by
\begin{equation}
    0 = \frac{\partial L}{\partial f}
    - \frac{\dd}{\dd \varphi} \left(\frac{\partial L}{\partial \left(\frac{\dd f}{\dd \varphi}\right)}\right),
\end{equation}
and with the definition in Eq. \eqref{particleTransformWithTime}, turns into the equation we wanted to prove:
\begin{equation}
    0 = \frac{\partial L}{\partial x} - \frac{\dd}{\dd t} \left(\frac{\partial L}{\partial \left(\frac{\dd x}{\dd t}\right)}\right)
    = \frac{\partial L}{\partial x} - \frac{\dd}{\dd t} \frac{\partial L}{\partial v}.
\end{equation}

\paragraph{Application to special relativity} \label{sectionApplicationToSpecialRelativity}
Let $\left(\begin{array}{c} \varphi \\ f \end{array} \right)$ be a Lorentz transformation between the two IRFs $T$ and $\bar{T}$.
Then, according to Eq. \eqref{derivationOfPhiWithRespectToTBar}, the transformation formula \eqref{definitionParticleLagrangianTransformationWithTime} reads
\begin{equation} \label{particleLagrangianLorentzTransformation}
\bar{L} \left(\bar{x}(\bar{t}), \frac{\dd \bar{x}}{\dd \bar{t}}, \bar{t} \right) = L \left(f, \frac{\dd f}{\dd \varphi},\varphi \right) \; \sqrt{1 - \frac{v^2}{c^2}} \;\; ,
\end{equation}
where $v$ is the relative velocity between $T$ and $\bar{T}$.

We next consider the case that $L$ fulfills the equation
\begin{align}
& L \left( \bar{x}(\bar{t}), \frac{\dd \bar{x}}{\dd \bar{t}}, \bar{t} \right) = L \left( f, \frac{\dd f}{\dd \varphi}, \varphi \right) \label{einsteinConditionRelativisticLagrangian} \\
\iff &
L \left( \bar{x}(\bar{t}), \frac{\dd \bar{x}}{\dd \bar{t}}, \bar{t} \right) = L \left( x(t), \frac{\dd x}{\dd t}, t \right) . \nonumber
\end{align}
A way to picture this condition is that $L$ is constructed in such a way that the transformation cancels out.
With this condition, Eq. \eqref{particleLagrangianLorentzTransformation} becomes
\begin{equation} \label{einsteinParticleLagrangianTransformation}
\bar{L} \left(\bar{x}(\bar{t}), \frac{\dd \bar{x}}{\dd \bar{t}}, \bar{t} \right)
= L \left(\bar{x}(\bar{t}), \frac{\dd \bar{x}}{\dd \bar{t}}, \bar{t} \right) \; \sqrt{1 - \frac{v^2}{c^2}} \;\; .
\end{equation}
The Euler-Lagrange equations in $\bar{T}$ and $T$ then read
\begin{align}
& \frac{\dd }{\dd \bar{t}} \frac{\partial}{\partial\frac{\dd \bar{x}}{\dd \bar{t}}} \bar{L} \left(\bar{x}(\bar{t}), \frac{\dd \bar{x}}{\dd \bar{t}}, \bar{t} \right)
- \frac{\partial}{\partial \bar{x}} \bar{L} \left(\bar{x}(\bar{t}), \frac{\dd \bar{x}}{\dd \bar{t}}, \bar{t} \right)  = 0  \nonumber \\
\iff \;\; &
 \frac{\dd }{\dd \bar{t}} \frac{\partial}{\partial\frac{\dd \bar{x}}{\dd \bar{t}}} \left( L \left(\bar{x}(\bar{t}), \frac{\dd \bar{x}}{\dd \bar{t}}, \bar{t} \right) \sqrt{1 - \frac{v^2}{c^2}} \right)
- \frac{\partial}{\partial \bar{x}} \left( L \left(\bar{x}(\bar{t}), \frac{\dd \bar{x}}{\dd \bar{t}}, \bar{t} \right) \sqrt{1 - \frac{v^2}{c^2}} \right) = 0 \label{eulerLagrangeRelParticleDashed}
\end{align}
and
\begin{align}
    & \frac{\dd }{\dd \varphi} \frac{\partial}{\partial\frac{\dd f}{\dd \varphi}} L \left(f, \frac{\dd f}{\dd \varphi}, \varphi \right)
    - \frac{\partial}{\partial f} L \left(f, \frac{\dd f}{\dd \varphi}, \varphi \right)  = 0  \nonumber \\
    \iff \;\; &
    \frac{\dd }{\dd \varphi} \frac{\partial}{\partial\frac{\dd f}{\dd \varphi}} \left(L \left(f, \frac{\dd f}{\dd \varphi}, \varphi \right) \underbrace{\sqrt{1 - \frac{0^2}{c^2}}}_{=1} \right)
    - \frac{\partial}{\partial f} \left(L \left(f, \frac{\dd f}{\dd \varphi}, \varphi \right) \underbrace{\sqrt{1 - \frac{0^2}{c^2}}}_{=1} \right)  = 0 . \label{eulerLagrangeRelParticleUndashed}
\end{align}
If we choose $T$ to be a particle's momentary rest IRF and $\bar{T}$ to be an observer IRF, then the two equations turn to
\begin{align}
    & \frac{\dd }{\dd \bar{t}} \frac{\partial}{\partial\frac{\dd \bar{x}}{\dd \bar{t}}} \left( L \left(\bar{x}(\bar{t}), \frac{\dd \bar{x}}{\dd \bar{t}}, \bar{t} \right) \sqrt{1 - \frac{\left(\frac{\dd \bar{x}}{\dd \bar{t}}\right)^2}{c^2}} \right)
    - \frac{\partial}{\partial \bar{x}} \left( L \left(\bar{x}(\bar{t}), \frac{\dd \bar{x}}{\dd \bar{t}}, \bar{t} \right) \sqrt{1 - \frac{\left(\frac{\dd \bar{x}}{\dd \bar{t}}\right)^2}{c^2}} \right) = 0 \nonumber \\
\iff \;\; &
    \frac{\dd }{\dd \bar{t}} \frac{\partial}{\partial \bar{v}} \left( L \left(\bar{x}, \bar{v}, \bar{t} \right) \sqrt{1 - \frac{\bar{v}^2}{c^2}} \right)
    - \frac{\partial}{\partial \bar{x}} \left( L \left(\bar{x}, \bar{v}, \bar{t} \right) \sqrt{1 - \frac{\bar{v}^2}{c^2}} \right) = 0 \label{eulerLagrangeRelParticleDashedFinal}
\end{align}
and
\begin{align}
   & \frac{\dd }{\dd \varphi} \frac{\partial}{\partial\frac{\dd f}{\dd \varphi}} \left(L \left(f, \frac{\dd f}{\dd \varphi}, \varphi \right) \underbrace{\sqrt{1 - \frac{0^2}{c^2}}}_{=1} \right)
    - \frac{\partial}{\partial f} \left(L \left(f, \frac{\dd f}{\dd \varphi}, \varphi \right) \underbrace{\sqrt{1 - \frac{0^2}{c^2}}}_{=1} \right)  = 0 \nonumber \\
    \iff \;\; &
    \frac{\dd }{\dd t} \frac{\partial}{\partial v} \left(L \left(x, v, t \right) \sqrt{1 - \frac{v^2}{c^2}} \right)
    - \frac{\partial}{\partial x} \left(L \left(x, v, t \right) \sqrt{1 - \frac{v^2}{c^2}} \right)  = 0, \label{eulerLagrangeRelParticleUndashedFinal}
\end{align}
where in Eq. \eqref{eulerLagrangeRelParticleUndashedFinal}, the velocity $v$ is zero as it is the particle's velocity in its momentary rest IRF.

The first postulate of special relativity requires the laws of physics to be the same in all IRFs.
The two equations \eqref{eulerLagrangeRelParticleDashedFinal} and \eqref{eulerLagrangeRelParticleUndashedFinal} obviously fulfill this postulate.
They may even be considered a mathematical formalization of the first postulate of special relativity.

From these considerations, we can read off the following rule for constructing particle Lagrangians of which the Euler-Lagrange equations fulfill the first postulate:
Find a function $L$ that fulfills equation \eqref{einsteinConditionRelativisticLagrangian} and multiply it by $\sqrt{1 - \frac{v^2}{c^2}}$.

\subsubsection{Heuristic argument} \label{sectionHeuristicFirstPostulate}
A common heuristic argument to derive the results of equations \eqref{eulerLagrangeRelParticleDashedFinal} and \eqref{eulerLagrangeRelParticleUndashedFinal} is to claim that, for special relativity, the action integral must be taken over proper time $\tau$ instead of classical time $t$:
\begin{equation} \label{relativisticAction}
    S = \int_{\tau_1}^{\tau_2} L \dd \tau.
\end{equation}
In an observer IRF with time $t$ and where the particle's velocity is $v=v(t)$, we may write Eq. \eqref{relativisticAction} in the form
\begin{equation} \label{relativisticActionWithObserver}
    S = \int_{\tau_1}^{\tau_2} L \dd \tau = \int_{t_1}^{t_2} L \; \sqrt{1-\frac{v^2}{c^2}} \; \dd t.
\end{equation}
Applying the principle of stationary action will then lead to Eq. \eqref{eulerLagrangeRelParticleUndashedFinal}.
The equations of motion will then be the same in any IRF if again we require Eq. \eqref{einsteinConditionRelativisticLagrangian} to hold.

\subsubsection{Einstein's original first postulate} \label{sectionEinsteinsOriginalFirstPostulate}
Einstein's original first postulate is different from the one we discussed in the end of section \ref{sectionApplicationToSpecialRelativity}.
In his first paper on relativity~\cite{EinsteinSpecialRelativity}, Einstein writes:
\\

``... suggest that the phenomena of electrodynamics as well as of mechanics possess no properties corresponding to the idea of absolute rest. They suggest rather that, as has already been shown to the first order of small quantities, the same laws of electrodynamics and optics will be valid for all frames of reference for which the equations of mechanics hold good. We will raise this conjecture (the purport
of which will hereafter be called the ``Principle of Relativity'') to the status
of a postulate ...''
\\

Thus, Einstein's original first postulate says that the laws of Maxwell's electrodynamics are the same in all IRFs \footnote{The original German version, found in Ref.~\cite{EinsteinSpecialRelativityOriginal}, reads: ``... f{\"u}hren zu der Vermutung, da{\ss} dem Begriffe der absoluten Ruhe nicht nur in der Mechanik, sondern auch in der Elektrodynamik keine Eigenschaften der Erscheinungen entsprechen, sondern da{\ss} vielmehr f{\"u}r alle Koordinatensysteme, f{\"u}r welche die mechanischen Gleichungen gelten, auch die gleichen elektrodynamischen und optischen Gesetze gelten, wie dies f{\"u}r die Gr{\"o}{\ss}en erster Ordnung bereits erwiesen ist. Wir wollen diese Vermutung (deren Inhalt im folgenden ``Prinzip der Relativit{\"a}t'' genannt werden wird) zur Voraussetzung erheben ...''}.
This means that Maxwell's equations as given in Ref.~\cite{WagnerGuthrieClassicalField} and the Lorentz force $F_L = e E + e v \times B$ are the same in all IRFs.

In the following section \ref{sectionConsequencesOfInvarianceLorentzForce}, we will assume that the Lorentz force law $F_L$ is the same in all IRFs.
However, we will not assume that Maxwell's equations are the same in all IRFs. 
Instead, we will be able to derive them using the results from section \ref{sectionConsequencesOfInvarianceLorentzForce}.
The original form of Einstein's first postulate will enter the discussion again in section \ref{sectionGauge}.

\subsubsection{Application to the Lorentz force law} \label{sectionConsequencesOfInvarianceLorentzForce}
According to Eq. \eqref{lorentzForceLagrangian}, the Lagrangian of the Lorentz force is given by
\begin{equation}
    L_L = - e (\phi - A \cdot v) = \frac{- e (\phi - A \cdot v)}{\sqrt{1-\frac{v^2}{c^2}}} \; \sqrt{1-\frac{v^2}{c^2}}.
\end{equation}
Section \ref{sectionApplicationToSpecialRelativity} tells us that the Lorentz force will be the same in two IRFs $\bar{T}$ and $T$
when
\begin{equation} \label{refomulatedLorenzForceLagrangian}
    \frac{- e (\bar{\phi} - \bar{A} \cdot \bar{v})}{\sqrt{1-\frac{\bar{v}^2}{c^2}}} = \frac{- e (\phi - A \cdot v)}{\sqrt{1-\frac{v^2}{c^2}}},
\end{equation}
where
\begin{itemize}
    \item $\bar{\phi}$ and $\bar{A}$ denote the electromagnetic potentials in $\bar{T}$,
    \item $\bar{v}$ denotes the particle's velocity in $\bar{T}$,
    \item $\phi$ and $A$ denote the electromagnetic potentials in $T$, and
    \item $v$ denotes the particle's velocity in $T$.
\end{itemize}
For the right hand side of Eq. \eqref{refomulatedLorenzForceLagrangian} we can write
\begin{align}
    \frac{- e \left(\phi - A \cdot v\right)}{\sqrt{1-\frac{v^2}{c^2}}}  &= - \frac{1}{\sqrt{1-\frac{v^2}{c^2}}} \; e \left(\frac{\phi}{c} , A_1, A_2, A_3\right) \; g
    \left(\begin{array}{c}
              c \\
              v_1\\
              v_2\\
              v_3\\
    \end{array} \right) \nonumber \\
    &= - e \left(\frac{\phi}{c} , A_1, A_2, A_3\right) \; g \; u, \label{invariantLorentzForceLagrangian2}
\end{align}
where in the last equation we used the 4-velocity $u$ from section \ref{section4Velocity}.
With that, Eq. \eqref{refomulatedLorenzForceLagrangian} can be written as
\begin{equation}
    - e \left(\frac{\bar{\phi}}{c} , \bar{A_1}, \bar{A_2}, \bar{A_3}\right) \; g \; \bar{u} = - e \left(\frac{\phi}{c} , A_1, A_2, A_3\right) \; g \; u.
\end{equation}
From section \ref{section4Velocity}, we know that $\bar{u}=\Lambda u$, with $\Lambda$ denoting the Lorentz transformation between $\bar{T}$ and $T$.
Thus, we find
\begin{equation}
- e \left(\frac{\bar{\phi}}{c} , \bar{A_1}, \bar{A_2}, \bar{A_3}\right) \; g \; \Lambda u = - e \left(\frac{\phi}{c} , A_1, A_2, A_3\right) \; g \; u .
\end{equation}
As this equation needs to hold for arbitrary 4-velocity $u$ we can write
\begin{equation}
\left(\frac{\bar{\phi}}{c} , \bar{A_1}, \bar{A_2}, \bar{A_3}\right) \; g \; \Lambda = \left(\frac{\phi}{c} , A_1, A_2, A_3\right) \; g.
\end{equation}
Because $g=\Lambda^\mathsf{T} g \Lambda$, the solution to this equation is
\begin{equation} \label{electromagenticPotentialsAre4VectorEquation}
    \left(\frac{\bar{\phi}}{c} , \bar{A_1}, \bar{A_2}, \bar{A_3}\right)
    = \left[ \Lambda
    \left(\begin{array}{c}
              \phi / c \\
              A_1\\
              A_2\\
              A_3\\
    \end{array} \right) \right]^\mathsf{T}
    \iff
    \left(\begin{array}{c}
              \bar{\phi} / c \\
              \bar{A_1}\\
              \bar{A_2}\\
              \bar{A_3}\\
    \end{array} \right)
    =
    \Lambda
    \left(\begin{array}{c}
              \phi / c \\
              A_1\\
              A_2\\
              A_3\\
    \end{array} \right),
\end{equation}
which has the important implication that the quantity
\begin{equation} \label{electromagenticPotentialsAre4Vector}
    \left(\begin{array}{c}
              \phi / c \\
              A_1\\
              A_2\\
              A_3\\
    \end{array} \right)
\end{equation}
is a 4-vector.

\subsubsection{The relativistic free particle} \label{sectionRelativisticFreeParticle}
Next, we are going to find the Lagrangian for a particle with zero net force acting upon it, also known as a free particle.

Section \ref{sectionConsequencesOfInvarianceLorentzForce}, relied heavily on the original form of Einstein's first postulate \ref{sectionEinsteinsOriginalFirstPostulate}. We cannot do so here simply because there is no electrodynamics involved.
What we are actually confronted with is to invent new physics, i.e. to invent a new Lagrangian.
The only premise we have is the rule from the end of section \ref{sectionApplicationToSpecialRelativity}.
At such a point, the usual approach is to apply the principle of Occam's razor and to make as few and as simple assumptions as can be thought of.

The simplest guess that can be thought of is that $L$ is some constant $k$ which is just the same in any IRF.
The way we will proceed is to compare this guess with the classical limit and see if it works.
In the case that we find no contradictions in that limit, we will find out the value of $k$.

Assume we observe the particle from within some observer IRF with time $t$.
In the observer IRF we assume the particle's velocity to be given by $v$.
The particle's Lagrangian in the observer IRF according to the rule from section \ref{sectionApplicationToSpecialRelativity} will then be
\begin{equation}
    L = k \sqrt{1-\frac{v^2}{c^2}}.
\end{equation}

To compare $k \sqrt{1 - \frac{v^2}{c^2}}$ to its classical limit, i.e. for the limit $v \ll c$, we first consider some mathematical preliminaries:\\

$(1+\epsilon)^\alpha$ for small $\epsilon$ can be approximated by the first two terms of its Taylor series:
\begin{equation}
    (1+\epsilon)^\alpha
    \approx (1+\epsilon)^\alpha \Big|_{\epsilon = 0}
    + \frac{\dd \; (1+\epsilon)^\alpha}{\dd \epsilon}\Big|_{\epsilon = 0} \epsilon
    = 1 + \alpha (1+\epsilon)^{\alpha -1} \Big|_{\epsilon = 0}  \epsilon
    = 1 + \alpha \epsilon.
\end{equation}

Applying this approximation to our relativistic Lagrangian with $\epsilon = - v^2/c^2$ results in

\begin{equation}
    L \approx k \left(1 + \frac{1}{2} \left(-\frac{v^2}{c^2} \right) \right)
    = k + \frac{1}{2} \left(-\frac{k}{c^2} \right) v^2.
\end{equation}

Since additional constants (in our case $k$) do not affect the equation of motion (i.e. the Euler-Lagrange equation) it is sufficient to compare the second term of this approximation to the free particle Lagrangian of classical mechanics, which is given by $L=\frac{1}{2} m v^2$.
The two become identical if we choose
\begin{equation}
    - \frac{k}{c^2} = m \iff k = -mc^2.
\end{equation}

Thus, our guess of the relativistic free particle Lagrangian is consistent with classical mechanics for small particle velocity $v$ if we write

\begin{equation} \label{freeRelativistivParticleLagrangian}
L = -mc^2 \sqrt{1-\frac{v^2}{c^2}}.
\end{equation}

\subsubsection{Energy of the relativistic free particle ($E=mc^2$)}

Using Eq. \eqref{legendre}, the energy of the free particle can be calculated from Eq. \eqref{freeRelativistivParticleLagrangian} as follows:

\begin{align}
    E &= \frac{\partial L}{\partial v} v - L \\
    &= -m c^2 \frac{1}{2 \sqrt{1-\frac{v^2}{c^2}}} \left(-\frac{2v}{c^2} \right) \cdot v
    -\left(-mc^2 \sqrt{1-\frac{v^2}{c^2}} \right) \\
    &= m c^2 \left(\frac{v^2/c^2}{\sqrt{1-\frac{v^2}{c^2}}} + \sqrt{1-\frac{v^2}{c^2}} \right) \\
    &= m c^2 \left(\frac{v^2/c^2 + 1 - v^2/c^2}{\sqrt{1-\frac{v^2}{c^2}}} \right) \\
    &= \frac{m c^2}{\sqrt{1-\frac{v^2}{c^2}}}. \label{einstein-famous-form}
\end{align}
For a particle at rest ($v=0$), Eq. \eqref{einstein-famous-form} takes Einstein's famous form, which tells us that in the theory of special relativity a particle with mass $m$ is assigned a \textit{rest energy}  $E = m c^2$.

\subsubsection{The equations of motion of the free relativistic particle} \label{sectionEquOfMotionFreeParticle}

From the result of Eq. \eqref{freeRelativistivParticleLagrangian}, the equations of motion for the free relativistic particle are given by

\begin{align}
    &0 = \frac{\dd}{\dd t} \frac{\partial }{\partial v_i} \left(-mc^2 \sqrt{1-\frac{v^2}{c^2}}\right) - \frac{\partial }{\partial x_i} \left(-mc^2 \sqrt{1-\frac{v^2}{c^2}}\right)
    \nonumber \\
    \iff &0 = \frac{\dd}{\dd t} \frac{\partial }{\partial v_i} \left(-mc^2 \sqrt{1-\frac{v^2}{c^2}}\right) \nonumber \\
    \iff &0 = \frac{\dd}{\dd t} \frac{\partial L}{\partial v_i} \;\; \text{for} \;\; i \in \{1,2,3\}.
\end{align}

We start with

\begin{equation}
    \frac{\partial L}{\partial v_i} = - m c^2 \frac{1}{2 \sqrt{1 - \frac{v^2}{c^2}}} \left(- \frac{2 v_i}{c^2}\right) = \frac{m v_i}{\sqrt{1 - \frac{v^2}{c^2}}}
\end{equation}

\begin{align}
   \implies \frac{\dd}{\dd t} \frac{\partial L}{\partial v_i}
   &= \frac{m \dot{v_i}}{\sqrt{1 - \frac{v^2}{c^2}}} + \frac{m v_i}{-2 \sqrt{1 - \frac{v^2}{c^2}}^3} \frac{-2 v_j \dot{v_j}}{c^2} \;\; \text{with implicit sum over} \;\; j \in \{1,2,3\} \nonumber \\
   &= \frac{m \dot{v_i}}{\sqrt{1 - \frac{v^2}{c^2}}} + \frac{m v_i}{\sqrt{1 - \frac{v^2}{c^2}}^3} \frac{v_j \dot{v_j}}{c^2} \nonumber \\
   &= \frac{m}{\sqrt{1 - \frac{v^2}{c^2}}} \left( \dot{v_i} + \frac{v_i}{\frac{c^2 - v^2}{c^2}} \frac{v_j \dot{v_j}}{c^2} \right)\nonumber \\
   &= \frac{m}{\sqrt{1 - \frac{v^2}{c^2}}} \left( \dot{v_i} + \frac{v_i}{c^2 - v^2} v_j \dot{v_j} \right)\nonumber \\
   &= \frac{m}{\sqrt{1 - \frac{v^2}{c^2}}} \frac{1}{c^2 -v^2} \left( (c^2 -v^2) \dot{v_i} + v_i v_j \dot{v_j} \right)\nonumber \\
   &= \frac{m}{\sqrt{1 - \frac{v^2}{c^2}}} \frac{1}{c^2 -v^2} \left( c^2 \dot{v_i} - v_j v_j \dot{v_i} + v_i v_j \dot{v_j} \right)\nonumber \\
   &= \frac{m}{\sqrt{1 - \frac{v^2}{c^2}}} \frac{c^2}{c^2 -v^2} \left( \dot{v_i} + \frac{v_i v_j \dot{v_j} - v_j v_j \dot{v_i}}{c^2} \right)\nonumber \\
   &= \frac{m}{\sqrt{1 - \frac{v^2}{c^2}}} \frac{c^2}{c^2\left(1- \frac{v^2}{c^2}\right)} \left( \dot{v_i} + \frac{v_i v_j \dot{v_j} - v_j v_j \dot{v_i}}{c^2} \right)\nonumber \\
   &= \frac{m}{\sqrt{1 - \frac{v^2}{c^2}}^3} \left( \dot{v_i} + \frac{v_i v_j \dot{v_j} - v_j v_j \dot{v_i}}{c^2} \right). \label{relativisticFreeEquationOfMotion}
\end{align}

The identity

\begin{align}
    [ v \times (v \times \dot{v})]_i &= \epsilon _{ijk} v_j (v \times \dot{v})_k \nonumber \\
    &= \epsilon _{ijk} v_j \epsilon _{kln} v_l \dot{v}_n \nonumber \\
    &= \epsilon_{kij} \epsilon _{kln}  v_j v_l \dot{v}_n \nonumber \\
    &= (\delta_{il} \delta_{jn} - \delta_{in} \delta_{jl})  v_j v_l \dot{v}_n \nonumber \\
    &= v_j v_i \dot{v}_j - v_j v_j \dot{v}_i \nonumber \\
\end{align}

allows us to rewrite the result of Eq. \eqref{relativisticFreeEquationOfMotion} as

\begin{equation}
    \frac{\dd}{\dd t} \frac{\partial L}{\partial v} = \frac{m}{\sqrt{1 - \frac{v^2}{c^2}}^3} \left( \dot{v} + \frac{1}{c^2} v \times (v \times \dot{v}) \right).
\end{equation}

As such, the equation of motion for the free relativistic particle reads

\begin{equation}
    0 = \frac{m}{\sqrt{1 - \frac{v^2}{c^2}}^3} \left(\dot{v} + \frac{1}{c^2} v \times (v \times \dot{v}) \right).
\end{equation}

\subsubsection{The Lagrangian and equations of motion of a relativistic particle in an electromagnetic field}

Using the results from sections \ref{sectionConsequencesOfInvarianceLorentzForce} and \ref{sectionRelativisticFreeParticle}, we can write down the Lagrangian of the relativistic particle  with charge $e$ in an electromagnetic field:
\begin{align} \label{lagrangianOfRelativisticParticleInEMField}
L &= -mc^2 \sqrt{1-\frac{v^2}{c^2}} \; - e (\phi - A \cdot v) \nonumber \\
&= \left( -mc^2 - e \left(\frac{\phi}{c} , A_1, A_2, A_3\right)  g  u\right) \sqrt{1-\frac{v^2}{c^2}}.
\end{align}

The equations of motion of the relativistic particle  with charge $e$ in an electromagnetic field
can be calculated from Eq. \eqref{lagrangianOfRelativisticParticleInEMField}, and with the results from sections \ref{sectionLorentzForceLagrangian} and \ref{sectionEquOfMotionFreeParticle}, reads:

\begin{align}
    F_L &= \frac{m}{\sqrt{1 - \frac{v^2}{c^2}}^3} \left( \dot{v} + \frac{1}{c^2} v \times ( v \times \dot{v}) \right) \nonumber \\
    \iff eE + e v \times B &= \frac{m}{\sqrt{1 - \frac{v^2}{c^2}}^3} \left( \dot{v} + \frac{1}{c^2} v \times (v \times \dot{v}) \right).
\end{align}


\section{Relativistic field Lagrangians} \label{sectionRelativisticFieldLagrangians}

In a previous paper~\cite{WagnerGuthrieClassicalField}, we derived the Lagrangian formalism for classical fields. 
The results there were classical in the sense that time was a special coordinate which was treated as being well-separated from spatial coordinates.
As we learned in section \ref{foundations}, in special relativity, spatial coordinates and time are not clearly separated.
This becomes obvious when we look at Eq. \eqref{lorentz}, where in contrast to Eq. \eqref{Galilei}, the spatial coordinate $x$ contributes to time.

It is for this reason that we wish to modify the Lagrangian formalism for fields in such a way that time and spatial coordinates are treated uniformly:

\begin{equation} \label{relativisticFielLagrangian}
    \mathcal{L} = \mathcal{L}\left(\psi,\frac{\partial \psi}{\partial q}\right),
\end{equation}
where $q$ denotes spatial coordinates including time and $\psi$ denotes the field.
The field may consist of multiple components.
A well-known example for a multiple component field is the electric field which, in classical non-relativistic electrodynamics, consists of three components that make up its direction and magnitude in space.

The most intuitive ansatz for an action $S$ created from $\mathcal{L}$ is

\begin{equation} \label{relativisticFielAction}
    S = \int_{A} \mathcal{L}\left(\psi,\frac{\partial \psi}{\partial q}\right) \dd q^{n},
\end{equation}
where $n$ denotes the number (dimension) of the coordinates $q$ and $A$ is an arbitrary $n$-dimensional area in the space of the coordinates $q$.
\\

As laid out in previous work ~\cite{WagnerGuthrie,WagnerGuthrieClassicalField}, we have clear criteria for deciding if the Lagrangian formalism arising from Eqs. \eqref{relativisticFielLagrangian} and \eqref{relativisticFielAction} is useful. These criteria are:

\begin{enumerate}
    \item Does the principle of stationary action lead to Euler-Lagrange equations?
    \item Are the Euler-Lagrange equations invariant under arbitrary differentiable and invertible transformations of the coordinates $q$ and the fields $\psi$?
    \item Does the Lagrangian $\mathcal{L}$ transform in a well-defined way?
\end{enumerate}
If these criteria are indeed fulfilled, we are well-motivated to find Lagrangians for physical field theories such that their field equations become the Euler-Lagrange equations of the Lagrangians.

\subsection{Euler-Lagrange equations} \label{sectionEulerLagrangeEquation}

For a detailed companion work to this section, see our previous paper in Ref.~\cite{WagnerGuthrieClassicalField}.
We consider the variations $\delta S$ of $S$ that result from variations $\delta \psi$ of the fields.
The variation of the fields is arbitrary except for the condition that it vanishes on the boundary of $A$ which we denote by $\partial A$:

\begin{equation}
    \delta \psi(q) = 0 \; \text{where} \; q \in \partial A.
\end{equation}
The variation of $S$ is given by

\begin{equation} \label{actionVariation}
    \delta S = \int_{A} \frac{\partial \mathcal{L}}{\partial \psi} \delta \psi
               + \frac{\partial \mathcal{L}}{\partial \frac{\partial \psi}{\partial q}} \;\; \delta \left(\frac{\partial \psi}{\partial q}\right) \dd q ^n.
\end{equation}
If we consider the possibly multidimensional components of $\psi$ indexed by $j$ and the $q$ coordinates by $i$ these summands mean:
\begin{equation}
    \frac{\partial \mathcal{L}}{\partial \psi} \delta \psi
    = \sum_{j} \frac{\partial \mathcal{L}}{\partial \psi_{j}} \; \delta \psi_{j}
\end{equation}
\begin{equation}
    \frac{\partial \mathcal{L}}{\partial \frac{\partial \psi}{\partial q}} \;\; \delta \left(\frac{\partial \psi} {\partial q}\right)
    = \sum_{i,j} \frac{\partial \mathcal{L}}{\partial \frac{\partial \psi_{j}}{\partial q_{i}}} \; \delta \left(\frac{\partial \psi_{j}} {\partial q_{i}}\right).
\end{equation}

We integrate the second summand in Eq. \eqref{actionVariation} by parts.
To do so we use the identity
$\delta \left(\frac{\partial \psi} {\partial q}\right)
= \frac{\partial \psi_2} {\partial q} - \frac{\partial \psi_1} {\partial q}
= \frac{\partial (\psi_2 - \psi_1)} {\partial q}
= \frac{\partial \delta \psi} {\partial q}$:

\begin{equation}
    \delta S = \int_{A}
    \frac{\partial \mathcal{L}}{\partial \psi} \delta \psi
    -\left(\frac{\partial}{\partial q} \cdot \left( \frac{\partial \mathcal{L}}{\partial \frac{\partial \psi}{\partial q}} \right)\right) \delta \psi 
    \dd q^n
    + \int_{A} \frac{\partial}{\partial q} \cdot \left( \frac{\partial \mathcal{L}}{\partial \frac{\partial \psi}{\partial q}} \cdot \delta \psi \right) \dd q^n.
\end{equation}
with ``$\frac{\partial}{\partial q} \cdot$'' we denote the divergence with respect to the coordinates $q$.
We refrain from using the usual ``$\nabla \cdot$'' because later the divergence with respect to other variables than $q$ will occur.

The second integral vanishes because of Gauss's theorem and $\delta \psi(q) = 0$ for any $q$ on the surface $\partial A$ of $A$.
\begin{equation}
\delta S = \int_{A}
\frac{\partial \mathcal{L}}{\partial \psi} \delta \psi
-\left(\frac{\partial}{\partial q} \cdot \left( \frac{\partial \mathcal{L}}{\partial \frac{\partial \psi}{\partial q}} \right)\right) \delta \psi 
\dd q^n.
\end{equation}
If we use the same index conventions for the field $\psi$ and $q$ as we did above, the last term means
\begin{equation}
    \left(\frac{\partial}{\partial q} \cdot \left( \frac{\partial \mathcal{L}}{\partial \frac{\partial \psi}{\partial q}} \right)\right) \delta \psi
    = \sum_j \left(\sum_i \frac{\partial}{\partial q_i} \left( \frac{\partial \mathcal{L}}{\partial \frac{\partial \psi_j}{\partial q_i}} \right)\right) \delta \psi_j.
\end{equation}
The last rewrite of $\delta S$ we perform is
\begin{equation}
    \delta S = \int_{A}
    \left(
    \frac{\partial \mathcal{L}}{\partial \psi}
    -\frac{\partial}{\partial q} \cdot \left( \frac{\partial \mathcal{L}}{\partial \frac{\partial \psi}{\partial q}} \right)\right) \delta \psi
    \dd q^n.
\end{equation}
Because the variation $\delta \psi$ is arbitrary (save for its endpoints), the only way to make $S$ stationary (which is equivalent to requiring $\delta S = 0$) is if $\mathcal{L}$ fulfills the condition
\begin{equation} \label{EulerLagrangeField}
    0 = \frac{\partial \mathcal{L}}{\partial \psi}
    -\frac{\partial}{\partial q} \cdot \left( \frac{\partial \mathcal{L}}{\partial \frac{\partial \psi}{\partial q}} \right).
\end{equation}
This is the Euler-Lagrange equation we were looking for.
Of course, this equation actually consists of multiple equations for the coordinates and the field components.
That is why it is common to use the plural and speak of the Euler-Lagrange equation\textit{\textbf{s}}.

\subsection{Invariance of the Euler-Lagrange equations under transformations} \label{LagrangeTranformation}

This section very closely follows section 3 in a previous work, available at Ref.~\cite{WagnerGuthrieClassicalField}.
Nonetheless, it is worth verifying that the arguments work without time as a special coordinate, too.
\\

Let $q=f(\bar{q})$ be an invertible and differentiable transformation of the coordinates and $\psi=F(\bar{\psi})$ be an invertible and differentiable transformation of the field.
We define the transformed Lagrangian  $\bar{\mathcal{L}}$ by

\begin{equation} \label{LagrTransform}
\bar{\mathcal{L}}\left(\bar{\psi}, \frac{\partial \bar{\psi}}{\partial \bar{q}}\right)
:= \mathcal{L}\left(F(\bar{\psi}) , \frac{\partial F(\bar{\psi})}{\partial f}\right)
\left| \mathrm{det} \frac{\partial f}{\partial \bar{q}} \right|
\end{equation}
where $\left| \mathrm{det} \frac{\partial f}{\partial \bar{q}} \right|$ is the absolute value of the determinant of the Jacobian matrix of $f$ with respect to the coordinates $\bar{q}$. \\

We will prove that, by requiring $S$ to be stationary, the two equations
\begin{equation} \label{ELGTransformed}
0 = \frac{\partial \bar{\mathcal{L}}}{\partial \bar{\psi}}
-\frac{\partial}{\partial \bar{q}} \cdot \left( \frac{\partial \mathcal{\bar{L}}}{\partial \frac{\partial \bar{\psi}}{\partial \bar{q}}} \right)
\end{equation}
and
\begin{equation} \label{ELGUntransformed}
0 = \frac{\partial \mathcal{L}}{\partial \psi}
-\frac{\partial}{\partial q} \cdot \left( \frac{\partial \mathcal{L}}{\partial \frac{\partial \psi}{\partial q}} \right)
\end{equation}
follow, and thus that the Euler-Lagrange equations are independent of arbitrary coordinate and field transformations, as long as the transformation of the Lagrangian is given by Eq. \eqref{LagrTransform}.
To do so, we consider arbitrary but small variations $\delta \bar{\psi}$ of the field $\bar{\psi}$ that vanish on the surface of an area of space $\bar{A}$.
We use this to find the condition for
\begin{equation}
    S = \int_{\bar{A}} \bar{\mathcal{L}}\left(\bar{\psi}, \frac{\partial \bar{\psi}}{\partial \bar{q}}\right) \dd \bar{q}^n
    = \int_{\bar{A}} \mathcal{L}\left(F(\bar{\psi}), \frac{\partial F(\bar{\psi})}{\partial f}\right)
    \left| \mathrm{det} \frac{\partial f}{\partial \bar{q}} \right| \dd \bar{q}^n
\end{equation}
to become stationary.

Equation \eqref{ELGTransformed} follows from repeating the considerations of section \ref{sectionEulerLagrangeEquation}. 
To prove Eq. \eqref{ELGUntransformed} we look at
\begin{equation}
    S = \int_{\bar{A}} \mathcal{L}\left(F(\bar{\psi}), \frac{\partial F(\bar{\psi})}{\partial f}\right)
    \left| \mathrm{det} \frac{\partial f}{\partial \bar{q}} \right| \dd \bar{q}^n,
\end{equation}
which by using the transformation formula of multidimensional integrals can be turned into
\begin{equation}
    S = \int_{f(\bar{A})} \mathcal{L}\left(F(\bar{\psi}), \frac{\partial F(\bar{\psi})}{\partial f}\right) \dd f^n,
\end{equation}
where $f(\bar{A})$ is the image of $\bar{A}$ under the coordinate transformation $f$. Based on this formula, the variation $\delta S$ of $S$ is given by
\begin{equation}
    \delta S = \int_{f(\bar{A})}
    \frac{\partial \mathcal{L}}{\partial F} \delta F
    + \frac{\partial \mathcal{L}}{\partial \frac{\partial F}{\partial f}}  \;\; \delta \left(\frac{\partial F} {\partial f}\right) \;\;
    \dd f^n,
\end{equation}
where
\begin{equation} \label{deltaFDefinition}
\delta F = \frac{\partial F}{\partial \bar{\psi}} \delta \bar{\psi}.
\end{equation}
Integration by parts of the second term leads to
\begin{equation} \label{calcDeltaSSection3}
\begin{split}
    \delta S = \int_{f(\bar{A})}
    \frac{\partial \mathcal{L}}{\partial F} \delta F
    -\left(\frac{\partial}{\partial f} \cdot \left( \frac{\partial \mathcal{L}}{\partial \frac{\partial F}{\partial f}} \right)\right) \delta F \;\;
    \dd f^n
    + \int_{f(\bar{A})} \frac{\partial}{\partial f} \cdot \left( \frac{\partial \mathcal{L}}{\partial \frac{\partial F}{\partial f}}  \delta F \right) \dd f^n,
\end{split}
\end{equation}
where the identity
$\delta \left(\frac{\partial F} {\partial f}\right)
= \frac{\partial F_2} {\partial f} - \frac{\partial F_1} {\partial f}
= \frac{\partial (F_2 - F_1)} {\partial f}
= \frac{\partial \delta F} {\partial f}$
was used.

The second integral in Eq. \eqref{calcDeltaSSection3} can be transformed into an integral over the surface of $f(\bar{A})$ which we denote by $\partial (f(\bar{A}))$. This surface is the same as the image of the surface of $\bar{A}$ under $f$:
\begin{equation}
    \partial (f(\bar{A})) = f(\partial \bar{A}).
\end{equation}
The simplest way to visualize this equation is to imagine a real area in space, which is described from within two systems of coordinates. To show that the third term vanishes, we will prove, that $\delta F$ is zero for any $q \in \partial (f(\bar{A}))$:

Let $q$ be an element of $\partial (f(\bar{A}))$ \footnote{The simplest way to picture this element is to imagine a real point on the surface of the area in space, which is described from within two systems of coordinates.}. 
Then for $q$ there exists a unique $\bar{q} \in \partial \bar{A}$ which is defined by $q=f(\bar{q})$.
We are going to use the fact from above that $\delta \bar{\psi}(\bar{q}) = 0$.
We recall that the variation $\delta \bar{\psi}$ is a difference between two fields which we name $\bar{\psi}_1$ and $\bar{\psi}_2$ such that
\begin{equation}
    \delta \bar{\psi} = \bar{\psi}_2 - \bar{\psi}_1.
\end{equation}
The value of $F$ considered as a function of $q$ is given by
\begin{equation}
    F(q) = F(\bar{\psi}(\bar{q})) \;\; \text{with} \;\; \bar{q} \;\; \text{defined through} \;\; q=f(\bar{q})
    \iff \bar{q} = f^{-1}(q). 
\end{equation}
The variation $\delta F$ that results from the difference $\delta \bar{\psi}$ between $\bar{\psi}_1$ and $\bar{\psi}_2$ is given by
\begin{equation}
    \delta F(q) = F(\bar{\psi}_2(\bar{q})) - F(\bar{\psi}_1(\bar{q}))
    = F(\bar{\psi}_1(\bar{q}) + \delta \bar{\psi} (\bar{q})) - F(\bar{\psi}_1(\bar{q}))
    = \frac{\partial F}{\partial \bar{\psi}} \delta \bar{\psi} (\bar{q}).
\end{equation}
Because $\delta \bar{\psi}(\bar{q})$ is zero by assumption, $\delta F(q)$ is zero, too, which finishes the proof.

\noindent As for $\delta S$, we are now left with
\begin{equation}
    \delta S = \int_{f(\bar{A})}
    \left(
    \frac{\partial \mathcal{L}}{\partial F}
    -\frac{\partial}{\partial f} \cdot \left( \frac{\partial \mathcal{L}}{\partial \frac{\partial F}{\partial f}} \right)\right) \delta F \;\;
    \dd q^n.
\end{equation}
As a consequence of Eq. \eqref{deltaFDefinition}, $\delta F$ is equally arbitrary as $\delta \bar{\psi}$. Thus the only way for $\delta S$ to become zero is if 
\begin{equation}
    0 =
    \frac{\partial \mathcal{L}}{\partial F}
    -\frac{\partial}{\partial f} \cdot \left( \frac{\partial \mathcal{L}}{\partial \frac{\partial F}{\partial f}} \right).
\end{equation}
If we now replace $F$ and $f$ according to their definitions by $\psi$ and $q$, this equation turns into Eq. \eqref{ELGUntransformed} and thus finishes the proof.

\subsection{Relativistic electrodynamics}

\subsubsection{Relativistic form of the Lagrangian of electrodynamics} \label{sectionRelativisticLagrangianElectrodynamics}
In literature on electrodynamics it is common to state that electrodynamics is a relativistic theory.
The most famous example is probably the passage of Einstein's original paper~\cite{EinsteinSpecialRelativityOriginal}.

Based on the previous two sections we will show that by two simple transformations of the nonrelativistic field Lagrangian of electrodynamics it can be made clear what this statement exactly means and in which sense it is true for Maxwell's field equations.
We start with the nonrelativistic field Lagrangian of electrodynamics from Ref.~\cite{WagnerGuthrieClassicalField}:
\begin{equation} \label{LagrangianElectDynClassical}
    \mathcal{L} = \epsilon_0 \frac{(-\nabla\phi - \frac{\partial A}{\partial t})^2 - c^2 (\nabla \times A)^2}{2} - \rho\phi + j \cdot A.
\end{equation}

In the same sense as section \ref{LagrangeTranformation}, we consider the following transformations:
\begin{itemize}
    \item time $t$ and the three spatial coordinates $x$;
    \begin{equation} \label{coordinateTransformClassical}
\left(\begin{array}{c}
          t
          \\
          x_1
          \\
          x_2
          \\
          x_3
\end{array} \right)
= f(x_0,x_1,x_2,x_3)
:=
\begin{pmatrix}
    1/c & 0 & 0 & 0
    \\
    0 & 1 & 0 & 0
    \\
    0 & 0 & 1 & 0
    \\
    0 & 0 & 0 & 1
\end{pmatrix}
\left(\begin{array}{c}
          x_0
          \\
          x_1
          \\
          x_2
          \\
          x_3
\end{array} \right)
=
\left(\begin{array}{c}
          x_0/c
          \\
          x_1
          \\
          x_2
          \\
          x_3
\end{array} \right)
\end{equation}
    \item the fields $\phi$ and $A$
    \begin{equation} \label{fieldTransformClassical}
    \left(\begin{array}{c}
              \phi
              \\
              A_1
              \\
              A_2
              \\
              A_3
    \end{array} \right)
    = F_A(A_0,A_1,A_2,A_3)
    :=
    \begin{pmatrix}
        c & 0 & 0 & 0
        \\
        0 & 1 & 0 & 0
        \\
        0 & 0 & 1 & 0
        \\
        0 & 0 & 0 & 1
    \end{pmatrix}
    \left(\begin{array}{c}
              A_0
              \\
              A_1
              \\
              A_2
              \\
              A_3
    \end{array} \right)
    =
    \left(\begin{array}{c}
              c A_0
              \\
              A_1
              \\
              A_2
              \\
              A_3
    \end{array} \right)
\end{equation}
    \item the charge density $\rho$ and the current density $j$;\footnote{
    One may argue that this transformation is nothing but a change of variables or a change of units and to treat it as a transformation of the Lagrangian is exaggerated.
    This true in the sense that the equations of motion (Maxwell's equations) can be rewritten in the new variables/units in a straightforward manner.
    Because this paper stresses the importance of the transformation behavior of the Lagrangian and the Euler-Lagrange equations, we found it adequate to use it at this point, too.
    Once the transformation is defined there is no discussion needed to apply it to the Lagrangian or to the equations of motion.
    All there is to is to apply rule (\ref{LagrTransform}), algebra, and calculus.
    Furthermore we think it's worth pointing out that a change of units can be interpreted as a coordinate transformation.
    As discussed in Ref.~\cite{WagnerGuthrie}, coordinates are not part of nature; they are a human means  to think about nature.
    Realizing that this holds for units too makes the argument relevant to everyday life, where units are quite inevitable.
    Nonetheless, because of their arbitrariness, even units cannot be part of reality itself.
    Those who like to philosophize may find that units are a suitable means to illustrate the shadows in Plato's Cave.
    The Lagrangian formalism restricts and makes clear the influence that coordinates and units have in scientific thinking.
}
    \begin{equation} \label{currentTransformClassical}
    \left(\begin{array}{c}
              \rho
              \\
              j_1
              \\
              j_2
              \\
              j_3
    \end{array} \right)
    = F_J (J_0,J_1,J_2,J_3)
    :=
    \begin{pmatrix}
        1/c & 0 & 0 & 0
        \\
        0 & 1 & 0 & 0
        \\
        0 & 0 & 1 & 0
        \\
        0 & 0 & 0 & 1
    \end{pmatrix}
    \left(\begin{array}{c}
              J_0
              \\
              J_1
              \\
              J_2
              \\
              J_3
    \end{array} \right)
    =
    \left(\begin{array}{c}
              J_0/c
              \\
              J_1
              \\
              J_2
              \\
              J_3
    \end{array} \right).
\end{equation}
\end{itemize}

The transformed Lagrangian, which we denote by $\mathcal{L}_R$, is given by

\begin{equation}
    \mathcal{L}_R = \frac{1}{c} \mathcal{L} \left(F, \frac{\partial F}{\partial f} \right),
\end{equation}
where  $F$ symbolizes $F_A$ and $F_J$.
The factor $1/c$ comes from the determinant of the matrix in Eq. \eqref{coordinateTransformClassical}.
This matrix is the Jacobian matrix of $f$ with respect to $x_0$,$x_1$,$x_2$, and $x_3$, and thus according to Eq. \eqref{LagrTransform}, the determinant of this matrix has to be included.

Replacing $\mathcal{L}$, $F$, and $f$ by their definitions results in

\begin{equation}
    \mathcal{L}_R = \frac{1}{c} \left\{ \frac{\epsilon_0}{2} \bigg[ \left(-c \nabla A_0 - \frac{\partial A}{\partial (\frac{x_0}{c})}\right)^2 - c^2 \left(\nabla \times A\right)^2 \bigg]
    - \left( J_0 A_0 - J \cdot A \right) \right\}
\end{equation}

\begin{equation} \label{maxwellHalfRelatifistic}
    = \frac{1}{c} \left\{ - \frac{c^2 \epsilon_0}{2} \bigg[ -\left(\frac{\partial A}{\partial x_0} + \nabla A_0 \right)^2 + \left(\nabla \times A\right)^2 \bigg]
    - \left( J_0 A_0 - J \cdot A \right) \right\},
\end{equation}
where $A$ and $J$ without index denote  $\left( \begin{array}{c} A_1 \\ A_2 \\ A_3 \end{array} \right)$ and $\left( \begin{array}{c} J_1 \\ J_2 \\ J_3 \end{array} \right)$ respectively.
For the next steps we will concentrate on the two terms in square brackets.
First we look at $\left(\frac{\partial A}{\partial x_0} + \nabla A_0 \right)^2$:

\begin{equation} \label{A0_Equation}
    \left(\frac{\partial A}{\partial x_0} + \nabla A_0 \right)^2
      = \left(\frac{\partial A_i}{\partial x_0} + \frac{\partial A_0}{\partial x_i} \right) \left(\frac{\partial A_i}{\partial x_0} + \frac{\partial A_0}{\partial x_i} \right),
\end{equation}
where we implicitly sum over the repeated index $i$ from $1$ to $3$.
\\
\\
\noindent We will now use the following definition of a new symbol $\partial$:

\begin{equation} \label{partialDerivTimesG}
\partial_0 := \frac{\partial}{\partial x_0} \;,\;
\partial_1 := -\frac{\partial}{\partial x_1} \;,\;
\partial_2 := -\frac{\partial}{\partial x_2} \;,\;
\partial_3 := -\frac{\partial}{\partial x_3}
\end{equation}
\\
\noindent
\textbf{Note:} The definition can also be written as:
\begin{equation}
    \left(\begin{array}{c}
              \partial_0
              \\
              \partial_1
              \\
              \partial_2
              \\
              \partial_3
    \end{array} \right)
    := g
    \left(\begin{array}{c}
              \frac{\partial}{\partial x_0}
              \\
              \frac{\partial}{\partial x_1}
              \\
              \frac{\partial}{\partial x_2}
              \\
              \frac{\partial}{\partial x_3}
    \end{array} \right)
    =
    \left(\begin{array}{c}
              \frac{\partial}{\partial x_0}
              \\
              -\frac{\partial}{\partial x_1}
              \\
              -\frac{\partial}{\partial x_2}
              \\
              -\frac{\partial}{\partial x_3}
    \end{array} \right)
\end{equation}
where the metric tensor $g$ was defined in Eq. \eqref{generalConditionForLorentzTransformations}.
\\
\\
\noindent
With definition \eqref{partialDerivTimesG}, Eq. \eqref{A0_Equation} becomes

\begin{equation}
\left(\frac{\partial A}{\partial x_0} + \nabla A_0 \right)^2
= \left(\partial_0 A_i - \partial_i A_0 \right) \left(\partial_0 A_i - \partial_i A_0 \right)
\end{equation}

\begin{equation}
= \frac{1}{2} \left[  \left(\partial_0 A_i - \partial_i A_0 \right) \left(\partial_0 A_i - \partial_i A_0 \right)
                    + \left(\partial_i A_0 - \partial_0 A_i \right) \left(\partial_i A_0 - \partial_0 A_i \right)\right],
\end{equation}

with $F_{0i} := \partial_0 A_i - \partial_i A_0 $ and $F_{i0} := \partial_i A_0 - \partial_0 A_i $,

\begin{equation}
   \left(\frac{\partial A}{\partial x_0} + \nabla A_0 \right)^2 = \frac{1}{2} \left[ F_{0i}F_{0i} + F_{i0}F_{i0} \right].
\end{equation}

Next we look at $(\nabla \times A)^2$:
\begin{equation}
    (\nabla \times A)^2 = \epsilon_{ijk} \epsilon_{ilm} \frac{\partial A_k}{\partial x_j} \frac{\partial A_m}{\partial x_l}
\end{equation}
where again we implicitly sum over all duplicate indices from $1$ to $3$.
\begin{equation}
    \iff (\nabla \times A)^2 = \epsilon_{ijk} \epsilon_{ilm} \partial_j A_k \partial_l A_m
\end{equation}
with the rule $\epsilon_{ijk} \epsilon_{ilm} = \delta_{jl}\delta_{km} - \delta_{jm}\delta_{kl}$ this can be written as
\begin{align}
    \iff (\nabla \times A)^2  
      &= (\delta_{jl}\delta_{km} - \delta_{jm}\delta_{kl}) \partial_j A_k \partial_l A_m \\
      &= \partial_j A_k \partial_j A_k - \partial_j A_k \partial_k A_j \\
      &= \frac{1}{2} \left[   \partial_j A_k \partial_j A_k - \partial_j A_k \partial_k A_j
                         + \underbrace{\partial_k A_j \partial_k A_j}_\text{$= \partial_j A_k \partial_j A_k$}  - \partial_j A_k \partial_k A_j \right] \\
      &= \frac{1}{2} \left[ (\partial_j A_k - \partial_k A_j) (\partial_j A_k - \partial_k A_j)  \right],
\end{align}
with $F_{jk} := (\partial_j A_k - \partial_k A_j)$,
\begin{equation}
    \iff (\nabla \times A)^2 = \frac{1}{2} F_{jk} F_{jk}.
\end{equation}

Thus, for the term $\left[ -\left(\frac{\partial A}{\partial x_0} + \nabla A_0 \right)^2 + \left(\nabla \times A\right)^2 \right]$
in Eq. \eqref{maxwellHalfRelatifistic}, we find:

\begin{equation}
    -\left(\frac{\partial A}{\partial x_0} + \nabla A_0 \right)^2 + \left(\nabla \times A\right)^2
    = \frac{1}{2} \left[ -(F_{0i} F_{0i} + F_{i0} F_{i0}) + F_{jk} F_{jk}\right].
\end{equation}

We define $F_{\mu\nu} := \partial_\mu A_\nu - \partial_\nu A_\mu$ with $\mu, \nu \in \{0,1,2,3\}$.
If we take into account that $F_{00} = 0$ follows from this definition, we can write

\begin{equation}
-\left(\frac{\partial A}{\partial x_0} + \nabla A_0 \right)^2 + \left(\nabla \times A\right)^2
= \frac{1}{2} F_{\mu\nu}F_{\alpha\beta} g_{\mu\alpha} g_{\nu\beta},
\end{equation}
where $g_{\mu\nu}$ are the elements of the metric tensor shown in Eq.  \eqref{generalConditionForLorentzTransformations}.
\\
\\
\noindent
\textbf{Note:} For the rest of this paper we will always consider Greek indices to run from $0$ to $3$ and will always implicitly sum over duplicate indices~\footnote{A note for experienced readers: In this paper we do not user upper and lower indices, we write metric tensors instead.}.
\\
\\
\noindent
We are now ready to put this result back into Eq. \eqref{maxwellHalfRelatifistic}:
\begin{equation}
    \mathcal{L}_R = \frac{1}{c} \left\{ - \frac{c^2 \epsilon_0}{2} \frac{1}{2} F_{\mu\nu}F_{\alpha\beta} g_{\mu\alpha} g_{\nu\beta} - \left( J_0 A_0 - J \cdot A \right) \right\},
\end{equation}
and with $1/\mu_0=c^2 \epsilon_0 $ this turns into
\begin{equation} \label{electrodynLagrangeRelativistic}
    \mathcal{L}_R = \frac{1}{c} \left\{ - \frac{1}{4 \mu_0} F_{\mu\nu}F_{\alpha\beta} g_{\mu\alpha} g_{\nu\beta} - J_\mu A_\nu g_{\mu\nu} \right\}.
\end{equation}

To pause for a moment of interpretation, notice that $\mathcal{L}_R$ in the form of Eq. \eqref{electrodynLagrangeRelativistic} is called the relativistic Lagrangian of electrodynamics.
The reason why it is called ``relativistic'' will be explained in the next section.
Similarly, equation \eqref{electrodynLagrangeRelativistic} is the result of the transformation given by Eqs. \eqref{coordinateTransformClassical}, \eqref{fieldTransformClassical}, and \eqref{currentTransformClassical}, as well as the application of the transformation rule \eqref{LagrTransform}.

\subsubsection{Why Maxwell's field equations are the same in every inertial reference frame} \label{sectionInvarianceOfElectroDynamicsLagrangian}

For $\mathcal{L}_R$ in the form of Eq. \eqref{electrodynLagrangeRelativistic}, we consider another transformation of the coordinates $x_0, x_1, x_2, x_3$, the fields $A_0,A_1,A_2,A_3$, and $J_0,J_1,J_2,J_3$:

\begin{equation} \label{coordinateTransform}
    x_\mu = f(\bar{x})_\mu := \Lambda_{\mu\nu} \bar{x}_\nu
\end{equation}

\begin{equation} \label{fieldTransform}
    A_\mu = F_A(\bar{A})_\mu := \Lambda_{\mu\nu} \bar{A}_\nu
\end{equation}

\begin{equation} \label{currentTransform}
    J_\mu = F_J(\bar{J})_\mu := \Lambda_{\mu\nu} \bar{J}_\nu,
\end{equation}
where $\Lambda$ is an arbitrary Lorentz transformation as discussed in section \ref{sectionNotations}.
In the above equations we used the same index-based notation as we did in the end of section \ref{sectionRelativisticLagrangianElectrodynamics}.
In the following, we will go on to use this notation.
Please be aware that, based on this notation, equation \eqref{generalConditionForLorentzTransformations} can be written as
\begin{equation} \label{invarianceWithMetricTensor}
    g_{\mu\nu} = \Lambda^\mathsf{T}_{\mu\alpha} g_{\alpha\beta} \Lambda_{\beta\nu}
    \; \iff \; g_{\mu\nu} = \Lambda_{\alpha\mu} g_{\alpha\beta} \Lambda_{\beta\nu}.
\end{equation}
\\
\noindent
\textbf{Note:}

\noindent
At this point, first and foremost, we are  interested in what happens to $\mathcal{L}_R$ when this transformation is applied using rule \eqref{LagrTransform}.
Strictly speaking, we are not required to be aware of the physical meaning of the transformation, but for readers who may not appreciate this abstraction it is helpful to note that
\begin{itemize}
    \item Eq. \eqref{coordinateTransform} is the way the coordinates transform in the real physical world.
    This was discussed in section \ref{sectionLorentzTransformation}.
    \item Eq. \eqref{fieldTransform} was shown in section \ref{sectionConsequencesOfInvarianceLorentzForce}.
    \item Eq. \eqref{currentTransform} is shown in appendix \ref{appendixConinuity}.
\end{itemize}
\noindent
First we look after the absolute value of the determinant of the Jacobian matrix in Eq. \eqref{LagrTransform}:
\begin{equation}
    \frac{\partial f}{\partial \bar{x}} = \Lambda \implies \left| \mathrm{det} \frac{\partial f}{\partial \bar{x}} \right| = |\mathrm{det} \Lambda|.
\end{equation}
From $g = \Lambda^\mathsf{T} g \Lambda$ follows the identity
\begin{equation} \label{determinantOfLorentzTransform}
    -1 = \mathrm{det} g = \mathrm{det} g \;  \mathrm{det}^2 \Lambda \; \implies \; 1 = \mathrm{det}^2 \Lambda \; \implies \; |\mathrm{det} \Lambda | = 1.
\end{equation}

To prepare for writing down $\bar{\mathcal{L}}_R$, we rewrite Eq. \eqref{electrodynLagrangeRelativistic} with $F_A$, $F_J$ and $\partial$ replaced by their definitions:
\begin{align}
    \mathcal{L}_R = & \frac{1}{c} \left\{ -\frac{1}{4\mu_0} \label{LagrangePartialA}
    \left(\partial_\mu A_\nu - \partial_\nu A_\mu \right)
    g_{\mu\alpha} g_{\nu\beta}
    \left(\partial_\alpha A_\beta - \partial_\beta A_\alpha\right) - J_\mu g_{\mu\nu} A_\nu \right\}
    \\
     = & \frac{1}{c} \bigg\{ -\frac{1}{4\mu_0} 
    \left(g_{\mu\tau} \frac{\partial A_\nu}{\partial x_\tau} - g_{\nu\epsilon} \frac{\partial A_\mu}{\partial x_\epsilon} \right)
    g_{\mu\alpha} g_{\nu\beta}
    \left(g_{\alpha\pi} \frac{\partial A_\beta}{\partial x_\pi} - g_{\beta\gamma} \frac{\partial A_\alpha}{\partial x_\gamma} \right) - J_\mu g_{\mu\nu} A_\nu \bigg\}.
\end{align}
By applying Eq. \eqref{LagrTransform} we find the transformed Lagrangian $\bar{\mathcal{L}}_R$ to be
\begin{align}
    \bar{\mathcal{L}}_R = & \frac{1}{c} \bigg\{ -\frac{1}{4\mu_0}  \nonumber \\
    & \left(g_{\mu\tau} \frac{\partial F_A(\bar{A})_\nu}{\partial f(\bar{x})_\tau} - g_{\nu\epsilon} \frac{\partial F_A(\bar{A})_\mu}{\partial f(\bar{x})_\epsilon} \right)
    g_{\mu\alpha} g_{\nu\beta}
    \left(g_{\alpha\pi} \frac{\partial F_A(\bar{A})_\beta}{\partial f(\bar{x})_\pi} - g_{\beta\gamma} \frac{\partial F_A(\bar{A})_\alpha}{\partial f(\bar{x})_\gamma} \right) \nonumber \\
    & - F_J(\bar{J})_\mu g_{\mu\nu} F_A(\bar{A})_\nu \bigg\}.
\end{align}

We now need to look for a way to express $g_{\mu\tau} \frac{\partial}{\partial f(\bar{x})_\tau}$ using the components of $\frac{\partial}{\partial \bar{x}}$.
To do so, we start with the chain rule:
\begin{equation}
    \frac{\partial}{\partial \bar{x}_\alpha} = \frac{\partial f_\nu}{\partial \bar{x}_\alpha} \frac{\partial}{\partial f_\nu}.
\end{equation}
Using the result from Eq. \eqref{coordinateTransform}, this turns into
\begin{equation}
    \frac{\partial}{\partial \bar{x}_\alpha} = \Lambda_{\nu\alpha} \frac{\partial}{\partial f_\nu}.
\end{equation}
Multiplying both sides with ${\Lambda_{\alpha\mu}}^{-1}$ gives
\begin{equation}
    {\Lambda_{\alpha\mu}}^{-1} \frac{\partial}{\partial \bar{x}_\alpha} = \frac{\partial}{\partial f_\nu} \delta_{\nu\mu} = \frac{\partial}{\partial f_\mu}.
\end{equation}
Multiplying both sides with $g_{\mu\beta}$ leads to
\begin{equation}
    {\Lambda_{\alpha\mu}}^{-1} g_{\mu\beta} \frac{\partial}{\partial \bar{x}_\alpha} = g_{\mu\beta} \frac{\partial}{\partial f_\mu}
    \iff (\Lambda^{-1}  g )_{\alpha\beta} \; \frac{\partial}{\partial \bar{x}_\alpha} = g_{\mu\beta} \frac{\partial}{\partial f_\mu}.
\end{equation}
Using $g = g^\mathsf{T}$ and $(\Lambda^{-1} g)^\mathsf{T} = g \Lambda^\mathsf{-1T}$ this turns into
\begin{equation}
    (g \Lambda^\mathsf{-1T} )_{\beta\alpha} \; \frac{\partial}{\partial \bar{x}_\alpha} = g_{\beta\mu} \frac{\partial}{\partial f_\mu}.
\end{equation}
By inverting both sides of Eq. \eqref{invarianceWithMetricTensor} and using $g=g^{-1}$, we find
$\Lambda^{-1}g\Lambda^\mathsf{-1T} = g \iff g\Lambda^\mathsf{-1T} = \Lambda g$.
Thus, we can write
\begin{equation}
    (\Lambda g )_{\beta\alpha} \; \frac{\partial}{\partial \bar{x}_\alpha} = g_{\beta\mu} \frac{\partial}{\partial f_\mu}.
\end{equation}
We use definition \eqref{partialDerivTimesG} for the transformed coordinates and write $\bar{\partial}_\nu := g_{\nu\mu} \frac{\partial}{\partial \bar{x}_\mu}$.
This leads us to
\begin{equation} \label{transformPartial}
    \Lambda_{\beta\nu} \bar{\partial}_\nu = g_{\beta\mu} \frac{\partial}{\partial f_\mu}
    \iff (\Lambda \bar{\partial})_\beta = g_{\beta\mu} \frac{\partial}{\partial f_\mu}.
\end{equation}
Although not needed in this section, it is worth noting that, with $x_\mu = f_\mu(\bar{x})$, we can give Eq. \eqref{transformPartial} the concise form
\begin{equation} \label{transformPartialConcise}
\Lambda \bar{\partial} = \partial,
\end{equation}
which shows that the symbol $\partial$ under Lorentz transformations transforms like a 4-vector.

We can now write $\bar{\mathcal{L}}_R$ as
\begin{align}
    \bar{\mathcal{L}}_R = \frac{1}{c} \bigg\{ & -\frac{1}{4\mu_0}  \nonumber \\
    & \left((\Lambda \bar{\partial})_\mu F_A(\bar{A})_\nu - (\Lambda \bar{\partial})_\nu F_A(\bar{A})_\mu \right)
    g_{\mu\alpha} g_{\nu\beta}
    \left((\Lambda \bar{\partial})_\alpha F_A(\bar{A})_\beta - (\Lambda \bar{\partial})_\beta F_A(\bar{A})_\alpha \right) \nonumber \\
    & - F_J(\bar{J})_\mu g_{\mu\nu} F_A(\bar{A})_\nu \bigg\}.
\end{align}
Next, we use Eqs. \eqref{fieldTransform} and \eqref{currentTransform}:
\begin{align} \label{LagrangeTransformed2}
    \bar{\mathcal{L}}_R = \frac{1}{c} \bigg\{ & -\frac{1}{4\mu_0}  \nonumber \\
    & \left((\Lambda \bar{\partial})_\mu (\Lambda\bar{A})_\nu - (\Lambda \bar{\partial})_\nu (\Lambda \bar{A})_\mu \right)
    g_{\mu\alpha} g_{\nu\beta}
    \left((\Lambda \bar{\partial})_\alpha (\Lambda \bar{A})_\beta - (\Lambda \bar{\partial})_\beta (\Lambda \bar{A})_\alpha \right) \nonumber \\
    & - (\Lambda \bar{J})_\mu g_{\mu\nu} (\Lambda \bar{A})_\nu \bigg\}.
\end{align}
Multiplying out the first summand, one of the resulting terms is
\begin{align}
    & (\Lambda \bar{\partial})_\mu (\Lambda\bar{A})_\nu g_{\mu\alpha} g_{\nu\beta} (\Lambda \bar{\partial})_\alpha (\Lambda \bar{A})_\beta \nonumber
    \\
    & =
      \Lambda_{\mu\epsilon} \bar{\partial}_\epsilon \Lambda_{\nu\tau} \bar{A}_\tau
      g_{\mu\alpha} g_{\nu\beta}
      \Lambda_{\alpha\pi} \bar{\partial}_\pi \Lambda_{\beta\sigma} \bar{A}_\sigma \nonumber
    \\
    & =
      \Lambda_{\mu\epsilon} g_{\mu\alpha} \Lambda_{\alpha\pi} \;
      \Lambda_{\nu\tau} g_{\nu\beta} \Lambda_{\beta\sigma} \;
      \bar{\partial}_\epsilon \bar{A}_\tau \bar{\partial}_\pi \bar{A}_\sigma \nonumber
    \\
      & =
      (\Lambda^\mathsf{T}g\Lambda)_{\epsilon\pi} \; (\Lambda^\mathsf{T}g\Lambda)_{\tau\sigma} \;
      \bar{\partial}_\epsilon \bar{A}_\tau \bar{\partial}_\pi \bar{A}_\sigma \nonumber
    \\
    & = g_{\epsilon\pi} \; g_{\tau\sigma} \; \bar{\partial}_\epsilon \bar{A}_\tau \bar{\partial}_\pi \bar{A}_\sigma.
\end{align}
Performing the following index transition $\epsilon \rightarrow \mu, \tau \rightarrow \nu, \pi \rightarrow \alpha, \sigma \rightarrow \beta$ turns this into
\begin{equation}
    g_{\mu\alpha} g_{\nu\beta} \bar{\partial}_\mu \bar{A}_\nu \bar{\partial}_\alpha \bar{A}_\beta.
\end{equation}
With analogous calculations for the remaining terms, Eq. \eqref{LagrangeTransformed2} becomes
\begin{equation} \label{LagrangeTransformFinal}
    \bar{\mathcal{L}}_R = \frac{1}{c} \left\{ -\frac{1}{4\mu_0}
    \left(\bar{\partial}_\mu \bar{A}_\nu - \bar{\partial}_\nu \bar{A}_\mu \right)
    g_{\mu\alpha} g_{\nu\beta}
    \left(\bar{\partial}_\alpha \bar{A}_\beta - \bar{\partial}_\beta \bar{A}_\alpha\right)
    - \bar{J}_\mu g_{\mu\nu} \bar{A}_\nu \right\}.
\end{equation}

\subsubsection{Interpretation} \label{sectionInvarianceMaxwell}

We found that the following equation holds: 
\begin{equation} \label{conditionForMaxwellLagriangian1}
    \bar{\mathcal{L}}_R \left( \bar{A}, \frac{\partial \bar{A}}{\partial \bar{x}} \right) = \mathcal{L}_R \left( \bar{A}, \frac{\partial \bar{A}}{\partial \bar{x}} \right).
\end{equation}
Using Eq. \eqref{LagrTransform} and the fact that $|\mathrm{det}\Lambda| = 1$ (see Eq. \eqref{determinantOfLorentzTransform}), we can also write
\begin{equation}\label{conditionForMaxwellLagriangian2}
    \bar{\mathcal{L}}_R \left( \bar{A}, \frac{\partial \bar{A}}{\partial \bar{x}} \right) = \mathcal{L}_R \left( A, \frac{\partial A}{\partial x} \right).
\end{equation}
This result together with the fact that the Euler-Lagrange equations are the same in all IRFs make Maxwell's equations fulfill the first postulate of special relativity which requires the laws of physics to be the same in all IRFs \footnote{Eq. (\ref{conditionForMaxwellLagriangian2}) is interesting to compare to Eq. (\ref{einsteinConditionRelativisticLagrangian}). One may say that in relativistic field theory $|\mathrm{det} \Lambda|$ plays the role of $\sqrt{1-\frac{v^2}{c^2}}$ in relativistic particle physics.}.
Thus, we arrive at the result mentioned in section \ref{sectionEinsteinsOriginalFirstPostulate}:
We derived that Maxwell's field equations are the same in all IRFs.

\subsubsection{Summary}
In section \ref{sectionApplicationToSpecialRelativity}, we found a rule to construct particle Lagrangians which fulfill the first postulate of special relativity.
Encouraged by Einstein's original first postulate cited in section \ref{sectionEinsteinsOriginalFirstPostulate}, we applied this rule to the Lagrangian of the Lorentz force in section \ref{sectionConsequencesOfInvarianceLorentzForce} and concluded that the electromagnetic potentials form a 4-vector. From the fact that the electromagnetic potentials form a 4-vector, we derived that Maxwell's equations fulfill the first postulate of special relativity, i.e. that they are the same in all IRFs.

All of these results were based on the Lagrangian formalism and its property that the Euler-Lagrange equations are the same in any coordinates. 
Although, to be precise, in section \ref{sectionProofEulerLagrangeRelPartcle1}, we found that there are limitations to the allowable coordinates in the case of particle physics and transformations which include time.

\section{Relativistic form of Maxwell's equations}
For the Lagrangian in Eq. \eqref{electrodynLagrangeRelativistic}, we calculate the equations of motion \eqref{EulerLagrangeField} of an arbitrary component $A_{\tau}$.
In this section we again make use of the definition $F_{\mu\nu} := \partial_\mu A_\nu - \partial_\nu A_\mu$ from section \ref{sectionRelativisticLagrangianElectrodynamics}.
We start with
\begin{align}
  \frac{\partial \mathcal{L}}{\partial\frac{\partial A_\tau}{\partial x_\epsilon}}
  = & \frac{\partial}{\partial\frac{\partial A_\tau}{\partial x_\epsilon}}
  \bigg[
  -\frac{1}{4 c \mu_0}
  \left(\partial_\mu A_\nu - \partial_\nu A_\mu \right)
  g_{\mu\alpha} g_{\nu\beta}
  \left(\partial_\alpha A_\beta - \partial_\beta A_\alpha\right)
  \bigg] \nonumber \\
  = & -\frac{1}{4 c \mu_0} \frac{\partial}{\partial\frac{\partial A_\tau}{\partial x_\epsilon}}
  \bigg[
  \left(g_{\mu\sigma} \frac{\partial A_\nu}{\partial x_\sigma} - g_{\nu\pi} \frac{\partial A_\mu}{\partial x_\pi} \right)
  g_{\mu\alpha} g_{\nu\beta}
  \left(g_{\alpha\eta} \frac{\partial A_\beta}{\partial x_\eta} - g_{\beta\gamma} \frac{\partial A_\alpha}{\partial x_\gamma} \right)
  \bigg] \nonumber \\ \nonumber \\
  = & -\frac{1}{4 c \mu_0}
  \bigg[
  \left(g_{\mu\sigma} \delta_{\tau\nu}\delta_{\epsilon\sigma} - g_{\nu\pi}\delta_{\mu\tau}\delta_{\epsilon\pi} \right)
  g_{\mu\alpha} g_{\nu\beta}
  \left(g_{\alpha\eta} \frac{\partial A_\beta}{\partial x_\eta} - g_{\beta\gamma} \frac{\partial A_\alpha}{\partial x_\gamma} \right) \nonumber \\
  & +  \left(g_{\mu\sigma} \frac{\partial A_\nu}{\partial x_\sigma} - g_{\nu\pi} \frac{\partial A_\mu}{\partial x_\pi} \right)
  g_{\mu\alpha} g_{\nu\beta}
  \left( g_{\alpha\eta}\delta_{\tau\beta}\delta_{\epsilon\eta} - g_{\beta\gamma}\delta_{\tau\alpha}\delta_{\epsilon\gamma} \right)
  \bigg] \nonumber \\ \nonumber \\
  = & -\frac{1}{4 c \mu_0}
  \bigg[
  \left(g_{\mu\epsilon} \delta_{\tau\nu} - g_{\nu\epsilon}\delta_{\mu\tau} \right)
  g_{\mu\alpha} g_{\nu\beta}
  \left(g_{\alpha\eta} \frac{\partial A_\beta}{\partial x_\eta} - g_{\beta\gamma} \frac{\partial A_\alpha}{\partial x_\gamma} \right) \nonumber \\
  & +  \left(g_{\mu\sigma} \frac{\partial A_\nu}{\partial x_\sigma} - g_{\nu\pi} \frac{\partial A_\mu}{\partial x_\pi} \right)
  g_{\mu\alpha} g_{\nu\beta}
  \left( g_{\alpha\epsilon}\delta_{\tau\beta} - g_{\beta\epsilon}\delta_{\tau\alpha} \right)
  \bigg] \nonumber \\ \nonumber \\
  = & -\frac{1}{4 c \mu_0}
  \bigg[
  \left(\delta_{\alpha\epsilon} g_{\tau\beta} - \delta_{\beta\epsilon}g_{\alpha\tau} \right)
  \left(g_{\alpha\eta} \frac{\partial A_\beta}{\partial x_\eta} - g_{\beta\gamma} \frac{\partial A_\alpha}{\partial x_\gamma} \right) \nonumber \\
  & + \left(g_{\mu\sigma} \frac{\partial A_\nu}{\partial x_\sigma} - g_{\nu\pi} \frac{\partial A_\mu}{\partial x_\pi} \right)
  \left( \delta_{\mu\epsilon}g_{\nu\tau} - \delta_{\nu\epsilon}g_{\mu\tau} \right)
  \bigg] \nonumber \\ \nonumber \\
  = & -\frac{1}{4 c \mu_0}
  \bigg[
  \left(
  g_{\epsilon\eta}g_{\tau\beta}\frac{\partial A_\beta}{\partial x_\eta} - \delta_{\tau\gamma}\frac{\partial A_\epsilon}{\partial x_\gamma}
  - \delta_{\tau\eta}\frac{\partial A_\epsilon}{\partial x_\eta} + g_{\epsilon\gamma}g_{\tau\alpha}\frac{\partial A_\alpha}{\partial x_\gamma}
  \right) \nonumber \\
  & + \left(
  g_{\epsilon\sigma}g_{\nu\tau}\frac{\partial A_\nu}{\partial x_\sigma} - \delta_{\sigma\tau}\frac{\partial A_\epsilon}{\partial x_\sigma}
  - \delta_{\pi\tau}\frac{\partial A_\epsilon}{\partial x_\pi} + g_{\epsilon\pi}g_{\mu\tau}\frac{\partial A_\mu}{\partial x_\pi}
  \right)
  \bigg] \nonumber \\ \nonumber \\
  = & -\frac{1}{4 c \mu_0}
  \bigg[
  g_{\epsilon\eta}g_{\tau\beta}\frac{\partial A_\beta}{\partial x_\eta} - \frac{\partial A_\epsilon}{\partial x_\tau}
  - \frac{\partial A_\epsilon}{\partial x_\tau} + g_{\epsilon\gamma}g_{\tau\alpha}\frac{\partial A_\alpha}{\partial x_\gamma} \nonumber \\
  & +   g_{\epsilon\sigma}g_{\nu\tau}\frac{\partial A_\nu}{\partial x_\sigma} - \frac{\partial A_\epsilon}{\partial x_\tau}
  - \frac{\partial A_\epsilon}{\partial x_\tau} + g_{\epsilon\pi}g_{\mu\tau}\frac{\partial A_\mu}{\partial x_\pi}
  \bigg] \nonumber
\end{align}
\begin{align}
   = & -\frac{1}{4 c \mu_0}
  \bigg[
  4 \cdot g_{\epsilon\eta}g_{\tau\beta}\frac{\partial A_\beta}{\partial x_\eta} - 4 \cdot \frac{\partial A_\epsilon}{\partial x_\tau}
  \bigg] \nonumber \\
  = & -\frac{1}{c \mu_0}
  \bigg[
  g_{\epsilon\eta}g_{\tau\beta}\frac{\partial A_\beta}{\partial x_\eta} - \frac{\partial A_\epsilon}{\partial x_\tau}
  \bigg] \nonumber \\
  = & -\frac{1}{c \mu_0}
  \bigg[
  g_{\tau\beta} \partial_\epsilon A_\beta - g_{\tau\beta}g_{\beta\alpha}\frac{\partial A_\epsilon}{\partial x_\alpha}
  \bigg] \nonumber \\
  = & -\frac{1}{c \mu_0} g_{\tau\beta}
  \bigg[
  \partial_\epsilon A_\beta - \partial_\beta A_\epsilon
  \bigg] \nonumber \\
  = & -\frac{1}{c \mu_0} g_{\tau\beta} F_{\epsilon\beta}. \nonumber
\end{align}
Next we look at
\begin{equation}
  \frac{\partial \mathcal{L}}{\partial A_\tau}
  = \frac{\partial }{\partial A_\tau} \left( - \frac{1}{c} J_\mu g_{\mu\nu} A_\nu \right)
  = - \frac{1}{c} J_\mu g_{\mu\nu} \delta_{\nu\tau}
  = - \frac{1}{c} J_\mu g_{\mu\tau}
  = - \frac{1}{c} g_{\tau\beta} J_\beta.
\end{equation}
Thus, according to Eq. \eqref{EulerLagrangeField}, the equation of motion is given by
\begin{align}
  0 &= - \frac{1}{c} g_{\tau\beta} J_\beta
  + \frac{\partial}{\partial x_\epsilon}
  \left(
  \frac{1}{c \mu_0} g_{\tau\beta}
  \left(
  \partial_\epsilon A_\beta - \partial_\beta A_\epsilon
  \right)
  \right) \\
  \iff 0 &= g_{\tau\beta} \left(J_\beta
  - \frac{1}{\mu_0}  \frac{\partial}{\partial x_\epsilon}\left( \partial_\epsilon A_\beta - \partial_\beta A_\epsilon \right) \right) \\
  \iff 0 &= J_\beta - \frac{1}{\mu_0}  \frac{\partial}{\partial x_\epsilon}\left( \partial_\epsilon A_\beta - \partial_\beta A_\epsilon \right) \\
  \iff \mu_0 J_\beta &= g_{\epsilon\alpha}g_{\alpha\pi}\frac{\partial}{\partial x_\pi}\left( \partial_\epsilon A_\beta - \partial_\beta A_\epsilon \right) \\
\iff \mu_0 J_\beta &= g_{\epsilon\alpha}  \partial_\alpha \left(\partial_\epsilon A_\beta - \partial_\beta A_\epsilon \right)
= g_{\epsilon\alpha}  \partial_\alpha F_{\epsilon\beta}. \label{equationOfMotionUntransformed}
\end{align}
This is the relativistic from of Maxwell's equations.
As mentioned in section \ref{sectionInvarianceMaxwell}, these equations fulfill the first postulate of special relativity, i.e. they are the same in all IRFs.

\section{Gauge and Lorentz transformations} \label{sectionGauge}
In this section, we will define gauge in the context of a particle Lagrangian and will explore how it is connected to gauge in the context of the electromagnetic potentials.
The aim of this section is to clarify the relationship between gauges of the electromagnetic potentials and the Lorentz transformation.

From section \ref{sectionConsequencesOfInvarianceLorentzForce}, we already know that the Lorentz transformation law for the potentials does not require any special gauge (i.e. the concept of gauge was not needed in section \ref{sectionConsequencesOfInvarianceLorentzForce} at all).
In section \ref{sectionEffectOfGaugeOnLorentzTransformations} we will show that a Lorentz transformation of the potentials cannot be achieved by a gauge transformation.
In section \ref{sectionTransformationPotentialsWithLorenzGauge} we will discuss an alternative proof of the Lorentz transformation law of the potentials which is less general than ours from section \ref{sectionConsequencesOfInvarianceLorentzForce}.
The loss of generality is due to the need to fix a gauge, namely the Lorenz (not Lorentz!) gauge.

\subsection{Gauge of a particle Lagrangian} \label{sectionGaugeParticleLagrangian}
A particle Lagrangian can be changed the following way without changing its equation of motion:

\begin{equation}
  L(q,\dot{q}, t) \rightarrow L(q,\dot{q}, t) + \frac{\dd}{ \dd t} F(q,t),
\end{equation}
where $F$ is an arbitrary function of the coordinates and time.
A change of this kind is called a gauge.

To prove that the equations of motion are not changed by a gauge we calculate the Euler-Lagrange equations for the right hand side:

\begin{align}
  & \frac{\dd}{\dd t} \frac{\partial \left(L + \frac{\dd F}{\dd t}\right)}{\partial \dot{q}} - \frac{\partial \left(L + \frac{\dd F}{\dd t}\right)}{\partial q} \nonumber \\
  & = \frac{\dd}{\dd t} \frac{\partial L}{\partial \dot{q}} - \frac{\partial L }{\partial q}
  + \frac{\dd}{\dd t} \frac{\partial}{\partial \dot{q}} \frac{\dd F}{\dd t}  - \frac{\partial}{\partial q} \frac{\dd F}{\dd t} \nonumber \\
  & = \frac{\dd}{\dd t} \frac{\partial L}{\partial \dot{q}} - \frac{\partial L }{\partial q}
  + \frac{\dd}{\dd t} \frac{\partial}{\partial \dot{q}} \left( \frac{\partial F}{\partial q} \dot{q} + \frac{\partial F}{\partial t} \right)
  - \frac{\partial}{\partial q} \left( \frac{\partial F}{\partial q} \dot{q} + \frac{\partial F}{\partial t} \right) \nonumber \\
  & = \frac{\dd}{\dd t} \frac{\partial L}{\partial \dot{q}} - \frac{\partial L }{\partial q}
  + \frac{\dd}{\dd t} \frac{\partial F}{\partial q}
  - \left(\frac{\partial^2 F}{\partial q^2} \dot{q} + \frac{\partial }{\partial t} \frac{\partial F}{\partial q} \right) \nonumber \\
  & = \frac{\dd}{\dd t} \frac{\partial L}{\partial \dot{q}} - \frac{\partial L }{\partial q}
  + \frac{\dd}{\dd t} \frac{\partial F}{\partial q}
  - \frac{\dd}{\dd t} \left(\frac{\partial F}{\partial q} \right) \nonumber \\
  & = \frac{\dd}{\dd t} \frac{\partial L}{\partial \dot{q}} - \frac{\partial L }{\partial q} + 0.
\end{align}
This is the same term as without gauge, which proves that the gauge does not change the equations of motion.

\subsection{Connection between the gauges of the particle Lagrangian and the electromagnetic field}
A gauge of the electromagnetic potentials is given by

\begin{align}
  A &\rightarrow A' = A + \nabla \lambda(x,t) \label{gaugeVectorPotential} \\
  \phi &\rightarrow \phi' = \phi - \frac{\partial}{\partial t} \lambda(x,t), \label{gaugeScalarPotential}
\end{align}
where $\lambda$ is an arbitrary function of the spatial coordinates and time.
This gauge is defined in such a way that it has no effect on the fields $E$ and $B$.
To prove this, we calculate the fields from the gauged potentials $A'$ and $\phi'$ using the formulas $B = \nabla \times A$ and $E = - \nabla \phi - \frac{\partial A}{\partial t}$, as defined for example in section 4 of Ref.~\cite{WagnerGuthrieClassicalField}:
\begin{align}
  B' &= \nabla \times A' = \nabla \times (A + \nabla \lambda) = \nabla \times A + 0 = \nabla \times A = B \\
  E' &= -\nabla \phi - \frac{\partial A'}{\partial t} = -\nabla \left( \phi - \frac{\partial \lambda}{\partial t} \right) -\frac{\partial}{\partial t} ( A + \nabla \lambda) \nonumber \\
  &= -\nabla \phi - \frac{\partial A}{\partial t} + \frac{\partial \nabla \lambda}{\partial t} - + \frac{\partial \nabla \lambda}{\partial t} = -\nabla \phi - \frac{\partial A}{\partial t} = E.
\end{align}

Next, we will calculate in which way this gauge will affect the Lagrangian $L_L$ of the Lorentz force:
the way $L_L$ changes by a gauge of the electromagnetic potentials is given by

\begin{align}
  L_L = - e (\phi - A \cdot v) \rightarrow & - e (\phi' - A' \cdot v) \nonumber \\
  & = -e\left(\phi-\frac{\partial \lambda}{\partial t}\right) - e ((A + \nabla \lambda) \cdot v) \nonumber \\
  & = - e (\phi - A \cdot v) - e \left(\frac{\partial \lambda}{\partial t} - \nabla \lambda \cdot v\right) \nonumber \\
  & = - e (\phi - A \cdot v) - e \frac{\dd \lambda}{\dd t} \nonumber \\
  & = - e (\phi - A \cdot v) - \frac{\dd  \left(e \lambda\right)}{\dd t}.
\end{align}
This shows that changing the gauge of the electromagnetic potentials changes the Lagrangian by a total time derivative.
As we saw in section \ref{sectionGaugeParticleLagrangian}, this change will not affect the equations of motion.
Both gauges are consistent to the extent that they do not change physical, i.e. measurable, phenomena.

\subsection{The effect of Lorentz transformations on physical phenomena in the case of a charged particle in an electromagnetic field} \label{sectionEffectOfGaugeOnLorentzTransformations}

We consider two IRFs $T, \bar{T}$ with coordinates $(t, x_1, x_2, x_3)$ and $(\bar{t}, \bar{x}_1, \bar{x}_2, \bar{x}_3)$ which move relative to each other with nonzero speed.
The equations of motion for a charged particle in an electromagnetic field in these two coordinate systems are given by
\begin{align}
  \frac{\dd}{\dd t} \frac{\partial L_{free}}{\partial v} - \frac{\partial L_{free}}{\partial x} &= e E + e v \times B \\
  \text{and} \nonumber \\
  \frac{\dd}{\dd \bar{t}} \frac{\partial \bar{L}_{free}}{\partial \bar{v}} - \frac{\partial \bar{L}_{free}}{\partial \bar{x}} &= e \bar{E} + e \bar{v} \times \bar{B},
\end{align}
where $L_{free}$ denotes the free particle Lagrangian $L_{free} = - m c^2 \sqrt{1 - \frac{v^2}{c^2}}$ \footnote{If the particle in $T$ and $\bar{T}$ moves with a speed small compared to the speed of light we may also choose $L_{free} = \frac{1}{2} m v^2$.}.
Although these equations are the same in all IRFs, the coordinates and velocity of the observed particle as well as the fields are certainly not the same in $T$ and $\bar{T}$ \footnote{Even $L_{free}$ is not equal to $\bar{L}_{free}$ because $\sqrt{1-v^2/c^2}$ and $\sqrt{1-\bar{v}^2/c^2}$ are not equal.}.

That these fields differ in $T$ and $\bar{T}$ is most famously discussed in the introduction of Einstein's first paper on special relativity for the example of a magnet and a conductor~\cite{EinsteinSpecialRelativity}:

``...if the magnet is in motion and the conductor at rest, there arises in the neighbourhood of the magnet an electric field with a certain definite energy, producing a current at the places where parts of the conductor are situated.
But if the magnet is stationary and the conductor in motion, no electric field arises in the neighbourhood of the magnet.
In the conductor, however, we find an electromotive force, to which in itself there is no corresponding energy, but which gives rise -- assuming equality of relative motion in the two cases discussed -- to electric currents of the same path and intensity as those produced by the electric forces in the former case.''

As changes in $E$ and $B$ result from Lorentz transformations of the potentials $\phi$ and $A$, it must be impossible to express a Lorentz transformation by a gauge transformation of $\phi$ and $A$.
This is because, by definition and construction, gauge transformations never change the fields.

\subsection{Lorenz gauge and the transformation law of the electromagnetic potentials} \label{sectionTransformationPotentialsWithLorenzGauge}
In this section we discuss a less general derivation of the quantity \eqref{electromagenticPotentialsAre4Vector}.
The derivation is less general because it requires a fixed gauge, specifically the Lorenz gauge.
We suspect that this derivation seduces some people who do not know the results of sections \ref{sectionConsequencesOfInvarianceLorentzForce} and \ref{sectionEffectOfGaugeOnLorentzTransformations} to believe in a stronger or different relation between Lorentz transformations and gauge transformations than there really is. 

The first step is to turn Eqs. \eqref{gaugeVectorPotential} and \eqref{gaugeScalarPotential} into a relativistic form using $\phi=c A_0, t=\frac{x_0}{c}, \partial_0 = \frac{\partial}{\partial x_0}, \partial_i = - \frac{\partial}{\partial x_i}$ for $i \in \{1,2,3\}$, see Eqs. \eqref{fieldTransformClassical} and \eqref{partialDerivTimesG}:
\begin{equation}
  A_\mu = \left(\begin{array}{c}
                      \phi /c
                      \\
                      A_1
                      \\
                      A_2
                      \\
                      A_3
          \end{array} \right)_\mu
  \rightarrow
  A'_\mu = \left(\begin{array}{c}
                                \phi' /c
                                \\
                                A'_1
                                \\
                                A'_2
                                \\
                                A'_3
  \end{array} \right)_\mu
  =
  \left(\begin{array}{c}
          \phi /c - \frac{1}{c} \frac{\partial \lambda}{\partial t}
          \\
          A_1 + \frac{\partial \lambda}{\partial x_1}
          \\
          A_2 + \frac{\partial \lambda}{\partial x_2}
          \\
          A_3 + \frac{\partial \lambda}{\partial x_3}
  \end{array} \right)_\mu
  =
  A_\mu - \partial_\mu \lambda.
\end{equation}
We now turn to Maxwell's equations in their relativistic form:
\begin{equation}
  \mu_0 J_\beta = g_{\epsilon\alpha}  \partial_\alpha \left(\partial_\epsilon A_\beta - \partial_\beta A_\epsilon \right)
  = g_{\epsilon\alpha} \partial_\alpha \partial_\epsilon A_\beta - g_{\epsilon\alpha} \partial_\alpha \partial_\beta A_\epsilon.
\end{equation}
The trick is now to choose the gauge function $\lambda$ in such a way that
\begin{equation}
  g_{\epsilon\alpha} \partial_\alpha \partial_\beta A_\epsilon = 0.
\end{equation}
This gauge is called the Lorenz gauge.
For the intrepid reader, the proof that such a gauge function exists can be found in Jackson's text on Electromagnetism~\cite[Sec. 3.2]{Jackson}.
In this gauge, Maxwell's equations take the simpler form
\begin{equation} \label{maxwellsEquationsLorenzGauged}
  \mu_0 J_\beta = g_{\epsilon\alpha} \partial_\alpha \partial_\epsilon A_\beta.
\end{equation}
To derive Eq. \eqref{electromagenticPotentialsAre4Vector} from Eq. \eqref{maxwellsEquationsLorenzGauged}, we consider two inertial reference frames $T$ and $\bar{T}$ with coordinates $x_\nu$ and $\bar{x}_\nu$, potentials $A_\nu$ and $\bar{A}_\nu$, and currents $J_\nu$ and $\bar{J}_\nu$. Using $\Lambda$, we denote the Lorentz transformation between the coordinates: $\bar{x}_\nu = \Lambda_{\nu \mu} x_\mu$.

According to Einstein's original first postulate \ref{sectionEinsteinsOriginalFirstPostulate}, Eq. \eqref{maxwellsEquationsLorenzGauged} in the coordinates of $\bar{T}$ is given by
\begin{equation} \label{maxwellsEquationsLorenzGaugedBarred}
  \mu_0 \bar{J}_\beta = g_{\epsilon\alpha} \bar{\partial}_\alpha \bar{\partial}_\epsilon \bar{A}_\beta.
\end{equation}
From appendix \ref{appendixConinuity} and Eq. \eqref{transformPartial}, we know that $\bar{J}_\nu = \Lambda_{\nu\gamma} J_\gamma$ and $\bar{\partial}_\nu = \Lambda_{\nu\gamma} \partial_\gamma$.
The substitution of these into Eq. \eqref{maxwellsEquationsLorenzGaugedBarred}, and using $\Lambda^\mathsf{T} g \Lambda = g$, leads to
\begin{equation}
  \mu_0 \Lambda_{\beta\gamma}J_\gamma = g_{\epsilon\alpha} \partial_\alpha \partial_\epsilon \bar{A}_\beta.
\end{equation}
The simplest way to make this equation consistent with equation \eqref{maxwellsEquationsLorenzGauged} is to assume
\begin{equation}
  \bar{A}_\beta = \Lambda_{\beta\gamma} A_\gamma,
\end{equation}
which is equivalent to Eq. \eqref{electromagenticPotentialsAre4VectorEquation}.


This proof is simpler than our proof from section \ref{sectionConsequencesOfInvarianceLorentzForce} because it requires a fixed gauge.
Another difference is that this proof assumes Einstein's original first postulate \ref{sectionEinsteinsOriginalFirstPostulate} to be true for Maxwell's field equations, while
in \ref{sectionConsequencesOfInvarianceLorentzForce} we assume Einstein's original first postulate to be true for the Lorentz force.

Looking back at section \ref{sectionInvarianceOfElectroDynamicsLagrangian}, the reader may be able to imagine that, with some effort, one could formulate another proof for the potentials being a 4-vector.
The steps/considerations to take would be:
\begin{itemize}
    \item Assume Maxwell's field equations are the same in all IRFs.
    \item Because the determinant of a Lorentz transformation  $\mathrm{det}(\Lambda)$ is $1$, the form invariance of Maxwell's equations is fulfilled, when the electrodynamic field Lagrangian fulfills Eqs. \eqref{conditionForMaxwellLagriangian1} and \eqref{conditionForMaxwellLagriangian2}.
    \item Write down the transformed field Lagranian and replace the coordinates and currents by their known transformations.
          Then conclude what would be the transformation of the potentials to make the Lagrangian invariant.
\end{itemize}

\subsection{Outlook on gauge field theories}
As with all topics of this paper, everything we considered about gauge until this section was known around the beginning of the last century.
After the second world war, gauge transformations began to play a new and surprising role in physics, which we feel should at least be mentioned.

When the Standard Model of Particle Physics~\cite{RevModPhys.71.S96} was developed, it was discovered that gauge transformations are generators of field theories, which describe the interaction between particles.
For example, in the Standard Model, the gauge transformation  of the electromagnetic potentials generates Maxwell's theory of  electrodynamics.
Field theories that are generated by gauge transformations are called ``gauge field theories.''
Other examples of gauge field theories are:
\begin{itemize}
    \item The theory of the electroweak interactions which unifies the theory of the electromagnetic field with those of the weak interaction. The fields of the weak interaction are the $Z^0$, the $W^+$, and the $W^-$ fields.
    \item The theory of the strong interaction, which describes the way quarks interact.
\end{itemize}
A most fascinating point is that, before the discovery of the gauge field theories, gauge transformations were just an aspect of Maxwell's theory.
An important contrast to this is that, in gauge field theories, gauge transformations become the main players and field theories, in the same way that Maxwell's theory are the implications thereof.

Interested readers may consult textbooks on Quantum Field Theory for further information.
With the exception of ``gauge field theory,'' the topic is often treated under the names ``Yang-Mills theory'' or ``Non-abelian gauge theory.''
We particularly recommend the following books for more detailed descriptions of these theories:
\begin{itemize}
    \item Zee~\cite{Zee2010} describes how electrodynamics emerges from gauge transformations at the beginning of chapter ``Non-abelian gauge theory.''
    \item Sterman~\cite{StermanChapter5_4} in the beginning of section ``Local gauge invariance'' stresses the importance of the topic in modern field theory.
    \item Schwartz~\cite{Schwartz2014} offers a detailed treatise, particularly in chapters ``Yang-Mills theory'' and ``Quantum Yang-Mills theory.''
\end{itemize}

Although it is important to keep in mind that these topics are increasingly active areas of research.

\appendix
\section{Transformation of $J_\mu$} \label{appendixConinuity}

This appendix, the results of which are mentioned in passing in Jackson~\cite[Sec. 11.9]{Jackson}, will motivate why Eq. \eqref{currentTransform} is the physically correct behavior of $J_\mu$ under Lorentz transformations.
We use the empirical fact that the electric charge is conserved.
The formula that describes this fact is the continuity equation:
\begin{equation}
    0 = \frac{\partial \rho}{\partial t} + \nabla \cdot j = \frac{\partial (c \rho)}{\partial (ct)} + \frac{\partial j_i}{\partial x_i}.
\end{equation}
with Eqs. \eqref{coordinateTransformClassical}, \eqref{currentTransformClassical}, and \eqref{partialDerivTimesG}, this equation can be written as
\begin{equation} \label{relativisticContinuity}
0 = \partial_0 J_0 - \partial_i J_i = g_{\mu\nu} \partial_\mu J_{\nu}.
\end{equation}
It is considered intuitively clear that charge is conserved in every inertial reference frame.
Thus, if we consider two inertial reference frames $\bar{T}$ and $T$  with coordinates $(\bar{x}_0, \bar{x}_1, \bar{x}_2, \bar{x}_3)$ and $(x_0, x_1, x_2, x_3)$
and currents $(\bar{J}_0, \bar{J}_1, \bar{J}_2, \bar{J}_3)$ and $(J_0, J_1, J_2, J_3)$,
we assume that the continuity equations in $\bar{T}$ and $T$ read
\begin{align}
  0 &= g_{\mu\nu} \bar{\partial}_\mu \bar{J}_{\nu}, \label{relativisticContinuityBarred} \\
  0 &= g_{\mu\nu} \partial_\mu J_{\nu}, \label{relativisticContinuityUnbarred}
\end{align}
respectively.
As both left sides are zero we can write
\begin{equation}
    g_{\mu\nu} \bar{\partial}_\mu \bar{J}_{\nu} = g_{\mu\nu} \partial_\mu J_{\nu}.
\end{equation}
Further, let $\Lambda$ be the Lorentz transformation, which transforms the coordinates of $\bar{T}$ into those of $T$. 
Then, along with Eq. \eqref{transformPartialConcise}, this becomes
\begin{equation} \label{currencyInvariance}
    g_{\mu\nu} \bar{\partial}_\mu \bar{J}_{\nu} = g_{\mu\nu} \Lambda_{\mu\alpha} \bar{\partial}_\alpha J_{\nu}.
\end{equation}
The solution to this equation is
\begin{equation} \label{currencyTransformationLaw}
J_\nu = \Lambda_{\nu\beta} \bar{J}_\beta,
\end{equation}
which allows us to rewrite Eq. \eqref{currencyInvariance} as
\begin{align}
g_{\mu\nu} \bar{\partial}_\mu \bar{J}_{\nu} &= g_{\mu\nu} \Lambda_{\mu\alpha} \bar{\partial}_\alpha \Lambda_{\nu\beta} \bar{J}_\beta \\
\iff g_{\mu\nu} \bar{\partial}_\mu \bar{J}_{\nu} &= \Lambda^{T}_{\alpha\mu} g_{\mu\nu} \Lambda_{\nu\beta} \;\; \bar{\partial}_\alpha \bar{J}_\beta \\
\iff g_{\mu\nu} \bar{\partial}_\mu \bar{J}_{\nu} &= (\Lambda^{T} g \Lambda)_{\alpha\beta} \;\; \bar{\partial}_\alpha \bar{J}_\beta \\
\iff g_{\mu\nu} \bar{\partial}_\mu \bar{J}_{\nu} &= g_{\mu\nu} \bar{\partial}_\mu \bar{J}_{\nu},
\end{align}
where in the last step we used Eq. \eqref{generalConditionForLorentzTransformations}.
The important result of this appendix is Eq. \eqref{currencyTransformationLaw}, which we used in Eq. \eqref{currentTransform}.

\bibliography{bibliography}













\end{document}